\documentclass[final,3p,times]{elsarticle}

\usepackage{graphicx}

\usepackage{amssymb}

\usepackage{amsmath}

\usepackage[dvips]{color}

\newcommand{\nn}{\nonumber}
\newcommand{\betaL}{\beta_{\mathrm{L}}}
\newcommand{\betaLE}{\beta_{\mathrm{L/E}}}
\newcommand{\betaLC}{\beta_{\mathrm{L}}^{~\mathrm{c}}}
\newcommand{\betaEC}{\beta_{\mathrm{E}}^{~\mathrm{c}}}
\newcommand{\betaLEC}{\beta_{\mathrm{L/E}}^{~\mathrm{c}}}

\newcommand{\gL}{g_{\mathrm{L}}}
\newcommand{\gE}{g_{\mathrm{E}}}
\newcommand{\gLE}{g_{\mathrm{L/E}}}
\newcommand{\gLC}{g_{\mathrm{L}}^{~\mathrm{c}}}

\newcommand{\gLR}{g_{\mathrm{L}}^{~\mathrm{ref}}}
\newcommand{\gTC}{g_{\mathrm{T}}^{~\mathrm{c}}}
\newcommand{\PBP}{a^3\langle\bar{\psi}\psi\rangle}
\newcommand{\LI}{\Lambda_{\mathrm{L}}^{\mathrm{imp}}}
\newcommand{\EI}{\Lambda_{\mathrm{E}}^{\mathrm{imp}}}
\newcommand{\LEI}{\Lambda_{\mathrm{L/E}}^{\mathrm{imp}}}
\newcommand{\gIRFP}{g_{\mathrm{4l}}^{\mathrm{IRFP}}}
\newcommand{\gSD}{g_{\mathrm{SD}}^{\mathrm{c}}}

\journal{Nuclear Physics B}

\begin{document}

\begin{frontmatter}



\title{Lattice Monte Carlo study of pre-conformal dynamics
in strongly flavoured QCD \\
in the light of the chiral phase transition at finite temperature}


\author[LNF]{Kohtaroh Miura}
\author[LNF]{Maria Paola Lombardo}

\address[LNF]{INFN Laboratori Nazionali di Frascati, 
	I-00044, Frascati (RM), Italy}

\begin{abstract}
We study the thermal phase transition 
in colour SU$(N_c=3)$ Quantum Chromodynamics (QCD)
with a variable number of fermions
in the fundamental representation by using
lattice Monte Carlo simulations.
We collect the (pseudo) critical couplings $\betaLC$
for $N_f=(0, 4, 6$, and $8$), 
and we investigate the pre-conformal dynamics
associated with the infra-red fixed point
in terms of the $N_f$ dependence
of the transition temperature.
{We propose three independent estimates of 
the number of flavour $N_f^*$ where the conformal
phase would emerge, which give consistent results within
the largish errors.  
We consider lines of fixed $N_t$ in the
space of ($N_f$, bare lattice coupling), and }
locate the vanishing of the step-scaling function
for $N_f^*\sim 11.1\pm 1.6$.
We define a typical interaction strength $\gTC$
at the scale of critical temperature $T_c$
and we find that $\gTC$ meets 
the zero temperature critical couplings
estimated by the two-loop Schwinger-Dyson equation
or the IRFP coupling in the four-loop beta-function
at $N_f^*\sim 12.5\pm 1.6$.
Further, we study the $N_f$ dependences of $T_c/M$
where $M$ is a UV $N_f$ independent reference scale
determined by utilising {the coupling at the
scale of the lattice spacing.} 
Then, $T_c/M$ turns out to be a decreasing function of $N_f$
and the vanishing $T_c/M$ indicates the emergence of the conformal window
at $N_f^* \sim 10.4 \pm 1.2$.
\end{abstract}


\end{frontmatter}




\section{Introduction}
{The analysis of the phases of strong interactions presents
many fascinating aspects --
mechanisms of  confinement,
 different realisations of the chiral symmetry,
the nature of the symmetric phase,
the emergence of  conformality, and many others.
All these topics are under active
scrutiny both theoretically and experimentally \cite{SCGT}.
While strong interactions spontaneously
break chiral symmetry in  ordinary QCD at zero temperature,
the chiral symmetry is realised either at
high temperatures -- in the so-called quark-gluon plasma (QGP)
phase -- and at a large number of flavours $N_f > N_f^*$
(even at zero temperature)
\cite{Appelquist:1996dq,Miransky:1997,Appelquist:1999hr,Sannino:2009za}.
In the latter case,
the theory is expected to become not only chirally but also 
conformally invariant. This is due to the emergence of
an infra-red fixed point (IRFP)
for $N_f > N_f^*$ at a coupling which is not 
strong enough to break  chiral symmetry. 
Both physics intuition and phenomenological analysis
based on functional renormalisation group~\cite{BraunGies}   
and finite temperature holographic 
QCD \cite{Alho:2012mh} 
indicate that the conformal phase of cold, many flavour QCD and 
the high temperature chirally symmetric phase are continuously connected. 
In particular, the onset of the conformal window coincides with 
the vanishing of the transition temperature, and the conformal
window appears as a zero temperature limit of
a possibly strongly interacting QGP.}

The analysis of the finite temperature phase transition
is a well-established
line of research within the lattice community,
and our approach will be completely conventional here.
According to the Pisarski-Wilczek 
scenario~\cite{Pisarski:1983ms}, the most likely possibility for $N_f \ge 3$
is a first order chiral transition in the chiral limit, turning into a 
crossover above a critical mass endpoint, and/or on lattices which are not
large enough. 
We will identify such crossover with confidence for a number of flavours
ranging from four to eight,
and we will complement these results with those
of the deconfinement transition in the quenched model. 
Then, we study the approach to the conformal phase
in the light of the chiral phase transition
at finite temperature with variable number of flavours.

One problem of this approach is the setting of 
a common scale among
theories which are essentially different.
We will propose two alternative
possibilities to handle this problem,
one evolving from our previous 
work \cite{Miura:2011mc}, and the other 
from a recent analysis \cite{Liao:2012tw}. 
Interestingly, this latter approach analyses the dependence of 
the confinement parameters on the matter content, and proposes
microscopic mechanisms for confinement
motivated by such $N_f$ dependence. 
Further, we will argue that
even results in the bare lattice parameters
can be used directly to locate
the critical number of flavours,
thus generalising to finite temperature
the Miransky-Yamawaki phase diagram, Ref. \cite{Miransky:1997}.

A second zero of the two-loop
beta-function of a non-Abelian gauge theory
implies, at least perturbatively,
the appearance of IRFP conformal symmetry
\cite{Caswell:1974gg,Banks:1981nn}.
In colour SU($3$) gauge
theory with $N_f$ massless fundamental fermions,
the second zero appears at $N_f\gtrsim 8.05$,
before the loss of asymptotic freedom (LAF) at
$N_f^{\mathrm{LAF}}=16.5$.
Analytic studies of the conformal transition of strong interactions
have produced a variety of predictions
for the conformal threshold:
the Schwinger-Dyson approach with
rainbow resummations
\cite{Appelquist:1996dq,Miransky:1997,Appelquist:1999hr}
or the functional renormalisation group method
\cite{BraunGies}  
suggest the onset of conformal window around $N_f^* \sim 12$.
An all-order perturbative beta-function
\cite{Ryttov:2007cx}
inspired by the NSVZ
beta-function of SQCD \cite{Novikov:1983uc} 
leads to a bound $N_f^* > 8.25$.
Instanton studies at large $N_f$ \cite{Velkovsky:1997fe}
claimed a qualitative change of behaviour at $N_f=6$.
$N_f^{*}$ has also been estimated
for different fermion representations \cite{Dietrich:2006cm}.

{The sub--critical region, 
when $N_f$ gets closer and closer to $N_f^*$, 
is interesting per se:
the question is whether the chiral dynamics
there shows any difference with the standard QCD dynamics.
Significant differences with respect to the QCD dynamics
might offer a basis to model builders
interested in beyond-the{--}standard{--}model theories. 
The recent discovery
of a 125 GeV boson at the LHC poses the question as to whether there
are light composite scalars which might be identified with such boson,
in alternative to a standard model Higgs boson.
Pre--conformal dynamics might well help these studies
~\cite{Sannino:2009za,Chivukula:2012ug}.
In our study, such pre-conformal dynamics could manifest itself 
either with a clear observation of a separation of scales,
or with a manifestation of a critical behaviour when approaching
$N_f^*$. One possibility is to observe the Miransky-Yamawaki essential 
singularity~\cite{Miransky:1997}.
Alternatively, in an FRG approach~\cite{BraunGies}, 
the pseudo-critical line is almost linear with $N_f$ 
for small $N_f$, and displays a singular behaviour when
approaching $N_f^*$, which could be the only observable
effects, beyond Miransky scaling.
A {\em jumping} scenario in 
which the change from a QCD dynamics to
the conformal window is abrupt is also a distinct possibility
\cite{Antipin:2012sm}.}

Clearly, as in any system undergoing a phase
transition, the nature and extent of the critical window are  purely
dynamical questions whose answer cannot be guessed a priori. 
Since  the underlying dynamics is completely non-perturbative, lattice
calculations are the only tool to perform an ab initio, rigorous
study of these phenomena, and many lattice studies
have recently appeared~\cite{DelDebbio:2010zz}.

This paper is one step of our
ongoing program \cite{Miura:xQCD12}--\cite{Deuzeman:2008sc}
which aims at elucidating the phase diagram of QCD
with fundamental fermions on the lattice, and in the continuum. 
Further studies 
either with fundamental fermions
\cite{Appelquist:2011dp} -- \cite{Fodor:2009wk}
or other representations 
\cite{Finland:MWT,Svetitsky:sextet,Kogut:2010cz,Fodor:2009ar,Fodor:2012ty}
have contributed to our current understanding of this
challenging field.
However, only a subset of these studies 
has addressed issues related with pre--conformal dynamics
\cite{Appelquist:2009ka}--
\cite{Lucini:MWT},
\cite{Miura:2011mc,Fodor:2012ni,Fodor:2012ty}
which are the main theme of this paper. 
{The direct inspection of theories at fixed $N_f$ is often inconclusive,
especially close to the expected threshold $N_f^*$. 
Also because of this, we feel it is a useful approach to try to
observe directly the approach to conformality by monitoring
the evolution of the pre--conformal results as a function of $N_f$.}


In this paper,
we investigate the thermal chiral phase transition
for $N_f=0,4,6,8$ colour SU$(N_c=3)$ QCD
by using lattice QCD Monte Carlo simulations
with staggered fermions.
Here, $N_f=6$ and $8$ is expected to be in the important regime
as suggested by the results
in Refs.~\cite{Velkovsky:1997fe,Appelquist:2011dp}.
We combine our findings with those of
our early work for $N_f=6$ and $8$
\cite{Miura:2011mc,Deuzeman:2008sc}.

This paper grows out of our early study 
\cite{Miura:2011mc} and extends it in several ways: 
We have accumulated more
statistics and added more parameters, and we present here
an extended set of simulations and details.
We develop a new scale setting procedure,
so that we can more confidently measure
the critical temperature on a common reference scale
among theories with different flavour content.
Furthermore, we present new estimates of the
critical number of flavours $N_f^*$.
Partly motivated by the recent work \cite{Liao:2012tw},
we introduce a typical interaction strength $\gTC$
at the critical temperature
based on our lattice results,
and compare it with a four-loop IRFP coupling
($\gIRFP$)~\cite{Ryttov:2012nt}
and a critical coupling ($g_{\mathrm{SD}}$) estimated by using
a two-loop Schwinger-Dyson equation~\cite{Appelquist:1998rb}.
Further, we introduce
and discuss the finite temperature version of the Miransky-Yamawaki
phase diagram, and propose a strategy to locate the critical
number of flavour motivated by the properties of the lattice
step-scaling function in the vicinity
of the IRFP \cite{Hasenfratz:2011xn}.
Some of the new results presented here have been anticipated in
a recent proceeding, and talks\cite{Miura:xQCD12,confx}.

This paper is organized as follows:
In the next section,
we will explain the simulation setups.
In Section \ref{sec:result},
we show our results for the chiral crossover at finite $T$,
for each $N_f$,
and then, we collect the critical lattice couplings
associated with the chiral crossovers at $N_f = 0,\ 4,\ 6,\ 8$.
In Section \ref{sec:AS},
we investigate the asymptotic scaling of
our critical couplings at each $N_f$.
In Section \ref{sec:discuss},
we investigate the $N_f$ dependences of the chiral crossovers,
and estimate the lower edge of the conformal window $N_f^*$.
Finally in Section \ref{sec:sum}, we provide concluding remarks.
The Appendix is devoted to the summary tables of
the simulation parameters and
the numerical results obtained by analysing the simulation outputs.

\section{Simulations' setup}\label{sec:setup}
We investigate  finite temperature QCD  with
different  number of flavour
$N_f = (0,\ 4,\ 6,\ 8)$
by utilising the  publicly available MILC code~\cite{MILC}.
The temperature $T$ is related to 
the inverse of the lattice temporal extension,
\begin{align}
&T\equiv \frac{1}{a(\beta_{\mathrm{L}})\cdot N_t}\ .\label{eq:T}
\end{align}
and we control it by varying $\betaL$ at fixed $N_t$.
The number of lattice points in
the spacial directions $N_s$ 
is  chosen such
that the aspect ratio $N_s/N_t \ge 2$ in all our runs.
For each $N_f$,
we use a single bare fermion mass $ma = 0.02$.
The simulation parameters
used in this study are summarised in \ref{app:sum_table}.

\begin{figure*}
\begin{center}
\includegraphics[width=6.5cm]{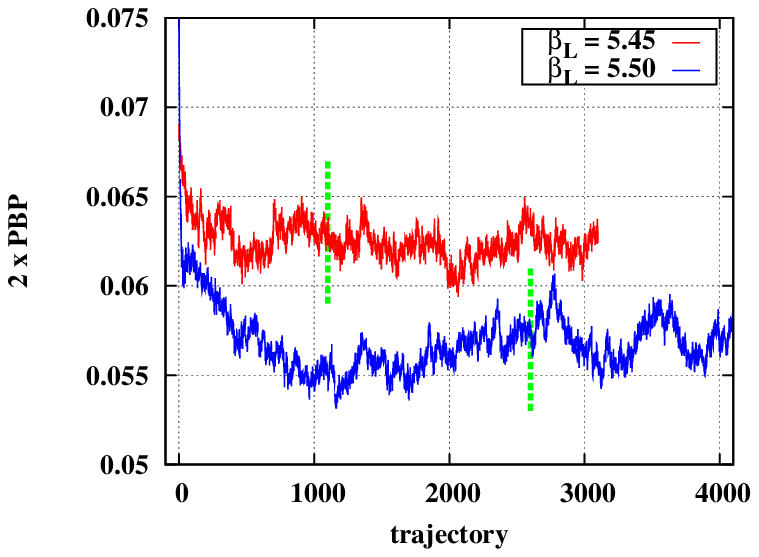}
\includegraphics[width=6.5cm]{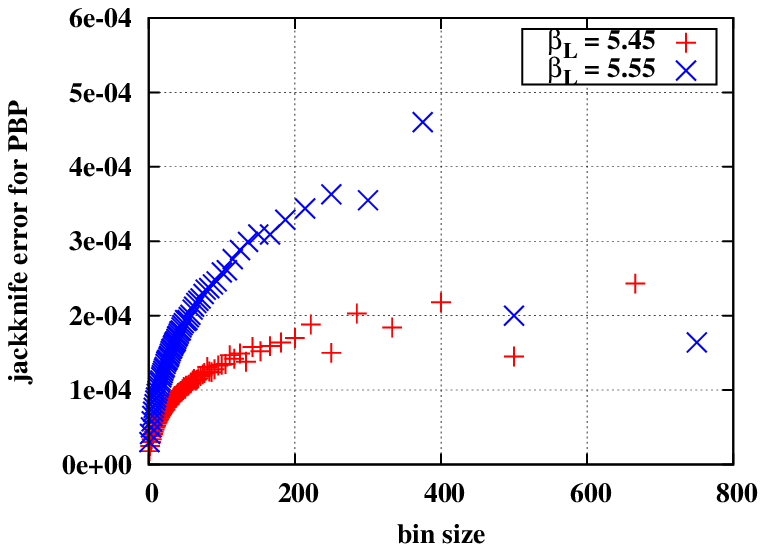}
\caption{
Left:
The Monte Carlo trajectories of the chiral condensate (PBP)
obtained by using the lattice volume $24^3\times 12$
just before the chiral crossover
$\betaL = 5.45 - 5.50$ at $N_f = 6$.
Right:
The jackknife errors as a function of a bin-size
for the trajectories shown in the left panel.
}
\label{Fig:traj}
\end{center}
\end{figure*}

{\subsection{Action and algorithm}}
The setup for the action explained below is the same
as the one used for $N_f=8$ in Ref.~\cite{Deuzeman:2008sc}
up to the number of flavour.
We use an improved version of the staggered action,
the Asqtad action~\cite{Lepage:1998vj},
with a one-loop Symanzik \cite{Bernard:2006nj,LuscherWeisz}
and tadpole~\cite{LM1985} improved gauge action,
\begin{align}
S = -\frac{N_f}{4}\mathrm{Tr}\log M[am,U,u_0]
+ \sum_{i=p,r,pg}\beta_i(g^2_{\mathrm{L}})
\mathrm{Re}\bigl[1-U_{C_i}\bigr]\ ,\label{eq:action}
\end{align}
where $\gL$ is the lattice bare coupling,
and $\beta_i$ are defined as
\begin{align}
&
\bigl(
\beta_p,\beta_r,\beta_{pg}
\bigr)
=
\biggl(
\frac{10}{\gL^2},
-\frac{\beta_p(1-0.4805\alpha_s)}{20u_0^2},
-\frac{\beta_p}{u_0^2}0.03325\alpha_s
\biggr)\ \label{eq:beta} \\
&
\alpha_s=-4\log\frac{u_0}{3.0684}\ ,\quad
u_0=\langle U_{C_p}\rangle^{1/4}\ .
\end{align}
The plaquette coupling
$\beta_p=10/\gL^2\equiv \beta_{\mathrm{L}}$
is a simulation input.
The $M[am,U,u_0]$ in Eq.~(\ref{eq:action})  denotes the matrix
for a single flavour Asqtad fermion with bare lattice mass $am$,
and $U_{C_i}$ represents the trace of
the ordered product of link variables along $C_i$,
for the $1\times 1$ plaquettes ($i=p$),
the $1\times 2$ and $2\times 1$ rectangles ($i=r$),
and the $1\times 1\times 1$ parallelograms ($i=pg$),
respectively -  all divided by the number of colours.
The tadpole factor $u_0$ is determined
by performing zero temperature simulations
on the $12^4$ lattice
(the second column of
Tables~\ref{tab:Nf0_Nt4} - \ref{tab:Nf8_Nt8}),
and used as an input
for finite temperature simulations.

To generate configurations with mass degenerate
dynamical flavours,
we have used the rational hybrid Monte Carlo algorithm
(RHMC)~\cite{Clark:2006wq}, which allows to 
simulate an arbitrary number of
flavours $N_f$ through varying the number of pseudo-fermions.
The quenched ($N_f = 0$) system has been realised
by using massive bare fermion mass $ma = 1.0$
in the four flavour system.
The six flavour system has been realised by using
two pseudo-fermions in the rational approximation
with a quarter root technique,
$N_f = 4\cdot 2\cdot 3/4 = 6$.
Then, we have assumed the rooting does not affect
the results within the accuracy of our simulation.
For the other number of flavour ($N_f = 0,\ 4,\ 8$),
we do not use the rooting.

We have adjusted the micro-canonical step length $\delta \tau$
and the step length of a single trajectory
$\Delta\tau=20\times \delta\tau$
to realise $75-80$ percent Metropolis acceptances.
Details are reported in the fourth column of
Tables~\ref{tab:Nf0_Nt4} - \ref{tab:Nf8_Nt8}.
For each parameter set,
we have collected a number of trajectories
ranging from a one thousand to ten thousands
- the latter closer to the chiral crossover regime.

\subsection{Observables}

The focus of this paper
is the analysis of the chiral phase transition
The fundamental observable is then the order parameter for
chiral symmetry, the chiral condensate:
\begin{equation}
a^3\langle\bar{\psi}\psi\rangle =
\frac{N_f}{4N_s^3N_t}
\Big\langle\mathrm{Tr\bigl[M^{-1}\bigr]}\Big\rangle
\ ,\label{eq:PBP}
\end{equation}
where $N_s~(N_t)$ represents the number of lattice sites
in the spatial (temporal) direction.
We have measured $a^3\langle\bar{\psi}\psi\rangle$
by using a stochastic estimator with 20 repetitions.
We have also measured
connected and disconnected chiral susceptibilities,
\begin{align}
a^2\chi_{\mathrm{conn}} &= 
-\frac{N_f}{4 N_s^3 N_t}
\langle \mathrm{Tr} \left[( MM )^{-1}\right ] \rangle
\ ,\nonumber\\
a^2\chi_{\mathrm{disc}} &=
\frac{N_f^2}{16 N_s^3 N_t}
\left [  \langle \mathrm{Tr} \left[M^{-1}\right] ^2\rangle
- \langle \mathrm{Tr} \left[M^{-1}\right] \rangle^2\right ]\ .
\end{align}
Here we have conveniently written the chiral condensate and its
susceptibilities in terms of traces of (products of) the staggered
fermion matrix $M$.
We note that the MILC convention for the chiral condensate
gives the twice of Eq.~(\ref{eq:PBP}), as will be indicated
several times in the following sections for results.
We have measured the susceptibilities $a^2\chi_{\mathrm{conn}}$
and $a^2\chi_{\mathrm{disc}}$ separately.

The disconnected chiral susceptibility
is a non-local quantity
which can be estimated from the variance of the bulk behaviour of
the chiral condensate.
Since we have used the stochastic estimator
for the chiral condensate measurements,
the variance would automatically include
part of the connected contributions
through random sources multiplying themselves.
Following Bernard et al. \cite{Bernard:1996zw},
we take into account  this effect
in our estimate for the disconnected part $a^2\chi_{\mathrm{disc}}$
by considering the only off-diagonal
elements of the covariance matrix for the random sources.

The measurements of $\PBP$ and $a^2\chi_{\mathrm{conn,diss}}$
allow us to construct two physically relevant quantities:
the scalar and pseudo-scalar susceptibilities,
\begin{align}
\chi_\sigma
&\equiv \frac {\partial \langle \bar \psi \psi\rangle}{\partial m}
= \chi_\mathrm{conn} + \chi_\mathrm{disc}\ ,\label{eq:sus_sig}\\
\chi_\pi 
&= \frac {\langle \bar \psi \psi\rangle}{m}\ .
\end{align}
Their associated cumulant
\begin{equation}
R_\pi \equiv \frac{\chi_\sigma}{\chi_\pi}\ ,
\label{eq:R_pi} 
\end{equation}
is a probe of the chiral symmetry
\cite{Deuzeman:2008sc,Kocic:1992is}.
This is owing to the fact that
$\chi_\sigma$ and $\chi_\pi$
are related through Ward identities to the spacetime volume
integral of the scalar ($\sigma$) and pseudo-scalar ($\pi$) propagators.
In the chiral limit,
the susceptibility ratio $R_{\pi}$ should be one
in chirally symmetric regime due to the degeneracy of the chiral partners,
while it should be zero in the spontaneously broken phase.
Even  including a finite bare fermion mass,
$R_{\pi}$ still has a strong signal for the chiral transition
or crossover. {In particular, $R_\pi \sim  1.0$ in the chirally symmetric
regime holds true till the chiral condensate is dominated 
by the linear mass term contribution.}
It turns out that $R_\pi$ allows the identification of 
a pseudo-critical coupling $\betaLC$
associated with the chiral crossover, 
{which , in the cases
we have studied, coincides in the error with the pseudo-critical
coupling determined from the maximum of the chiral susceptibility.}

In the gauge sector,
we measure the Polyakov loop,
\begin{equation}
L =
\frac{1}{N_cN_s^3}\sum_{\mathbf{x}}
\mathrm{Re}
\bigg\langle
\mathrm{tr}_c\prod_{t=1}^{N_t}U_{4,t\mathbf{x}}
\bigg\rangle
\ ,\label{eq:PLOOP}
\end{equation}
where $\mathrm{tr}_c$ denotes the trace in colour space,
and $U_{4,t\mathbf{x}}$ is the temporal link variable.
From the variance of $L$,
we also evaluate the susceptibility for the Polyakov loops.

\subsection{Statistics, and error analysis}
In the vicinity of the chiral crossover,
we have a long auto-correlation time,
{and {thermalization} checks require extra care.} 
Here we explain our analyses by using a typical example:
The left panel of Fig. \ref{Fig:traj} displays
the  evolution of the chiral condensate 
on the lattice volume $24^3\times 12$
just before the chiral crossover
$\betaL = 5.45$  and $5.50$ at $N_f = 6$,
one of the most time-consuming examples
in our simulations.
(In order to shorten the simulation time,
we started the evolution from 
thermalized configurations obtained at $\betaL < 5.45$.)
We have computed the ensemble averages 
by  using the 
last $2500$ ($2000$) trajectories at $\betaL = 5.45$ ($5.50$)
and we have confirmed that they 
are consistent with those obtained by using
last $2000$ ($1500$) trajectories. We have then 
used  the latter trajectories to evaluate the average.
In the cases we are considering, 
this corresponds to the data found
in the right-hand side of the vertical green (dashed) lines
in the left panel of Fig. \ref{Fig:traj}.

We divide the obtained data set into several bins
and utilise the jackknife method
in order to take into account 
the auto-correlation effect in the error estimate.
As a bin-size $s_{\mathrm{bin}}$ becomes larger,
the jackknife error increases 
(the right panel of Fig.~\ref{Fig:traj}),
which is due to the decrease of
the effective number of (uncorrelated) data
($n_{\mathrm{ave}}/s_{\mathrm{bin}}$,
$n_{\mathrm{ave}} = $
the number of trajectories to calculate the average).
For a sufficiently large $s_{\mathrm{bin}}$,
the jackknife errors at $\betaL = 5.45$ and $5.50$
level off, giving a reliable error estimate.

Here is the result obtained from the above procedures:
\begin{align}
\betaL &\qquad\qquad
n_{\mathrm{traj}} &
n_{\mathrm{ave}} &\qquad\qquad
s_{\mathrm{bin}} &
2\PBP \nn\\
5.45 &\qquad\qquad
3100 &
2000 &\qquad\qquad
400 &
0.0622(2)\ , \nn\\
5.50 &\qquad\qquad
4100 &
1500 &\qquad\qquad
375 &
0.0570(5)\ , \nn
\end{align}
We have performed the analyses explained here
for all the various $\betaL$, $N_f$, and the lattice volumes.
The results are summarised in \ref{app:sum_table}.

\section{Results on the lattice thermal transition}\label{sec:result}
In this section,
we show our simulation results
on the chiral and deconfinement crossover 
for the different number of flavours $N_f$.

\begin{table*}
\caption{
Summary of the (pseudo) critical lattice couplings $\betaLC$
for the theories with $N_f=0,~4,~6,~8$, $am=0.02$
and varying $N_t=4,~6,~8,~12$.
The entries with $\ast$
are the update for our previous results \protect\cite{Miura:2011mc}.
The entries with $\dagger$
have been quoted from our previous studies on $N_f = 8$
\protect\cite{Deuzeman:2008sc}.
}\label{Tab:bc}
\begin{center}
\begin{tabular}{c|cccc}
\hline\hline
$N_f\backslash N_t$ &	
$4$&
$6$&
$8$&
$12$\\
\hline
$0$ &
$7.35\pm 0.05$&		
$7.97^{\ast}\pm 0.07$&	
$8.26\pm 0.06$&		
$-$\\
$4$ &		
$5.65\pm 0.05$&		
$6.00^{\ast}\pm 0.05$&	
$6.15\pm 0.15$&		
$-$\\
$6$ &
$4.675^{\ast}\pm 0.05$&	
$5.025^{\ast}\pm 0.05$&	
$5.20^{\ast}\pm 0.05$&	
$5.55^{\ast}\pm 0.1$\\	
$8$ &
$-$&
$4.1125^{\dagger}\pm 0.0125$&
$4.275\pm 0.05$&
$4.34^{\dagger}\pm 0.04$\\
\hline\hline
\end{tabular}
\end{center}
\end{table*}

We have used a common bare fermion mass $ma = 0.02$
for all simulations at finite $N_f$.
According to the Pisarski-Wilczek scenario~\cite{Pisarski:1983ms},
the most likely possibility for $N_f \ge 3$
is a first order chiral transition in the chiral limit.
Introducing a bare fermion mass,
the first order phase transition
will eventually turn into a crossover
for masses larger than some  critical mass.
Since the chiral condensate looks smooth in our results,
we are most likely above the critical endpoint in all 
the cases we have studied, and 
we use the terminology of ``chiral crossover'' in the following.

The finite bare mass $ma=0.02$
might have a different physical relevance at each $N_f$,
as well as for different bare coupling for a fixed $N_f$.
It remains then to be seen how our results would change in the
chiral limit, and we hope to come back to this
point in a future study. Since we have noted that at strong coupling
and small masses the improvement term in the Action might 
be responsible for the spurious phases \cite{Deuzeman:2012ee,daSilva:2012wg} 
observed also in Ref. \cite{Cheng:2011ic,Deuzeman:2011pa}, 
we might also consider an unimproved action for this study.

Before entering into details, let us
summarise our main results, {\em i.e.}
the critical lattice couplings $\betaLC$
associated with the chiral crossover in Table \ref{Tab:bc}.
For $N_f = (4,6,8)$ 
we have observed that 
the peak position of the chiral susceptibility $a^2\chi_{\sigma}$,
whenever clearly defined,
coincides within the
errors with the inflection point
of $R_{\pi}$ defined in Eq.~(\ref{eq:R_pi}), as well
as with the inflection point of the chiral condensate
and that of the Polyakov loop. 
{This indicates that the crossover region is rather narrow,
as different indicators give consistent pseudo-critical points.
We then quote the  common pseudo-scalar coupling, 
with a conservative error estimate.}
For the quenched ($N_f = 0$) case,
we have extracted the pseudo-critical coupling
from the deconfinement crossover by evaluating
the peak position of the Polyakov loop susceptibility.

In the following subsections,
we present these results in detail,
starting from $N_f = 6$ and $8$
in the first two subsections, and continuing with the
$N_f = 4$ and $N_f=0$. 
The reader who is not interested in these technical
details is advised to skip the rest of this Section and proceed
directly to the next one.

\subsection{Chiral crossover at $N_f = 6$}
\label{subsec:Nf6}
\begin{figure*}
\begin{center}
\includegraphics[width=6.5cm]{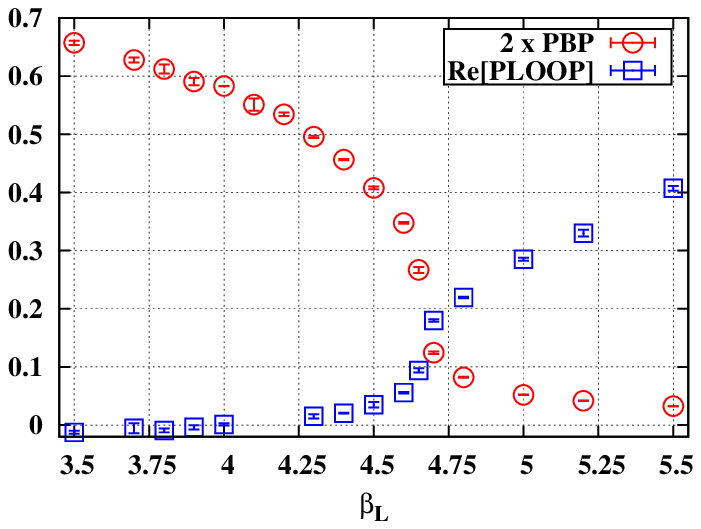}
\includegraphics[width=6.5cm]{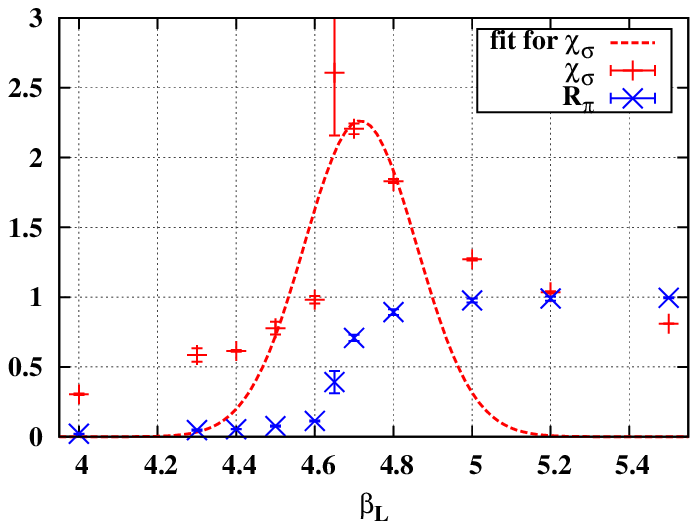}

\includegraphics[width=6.5cm]{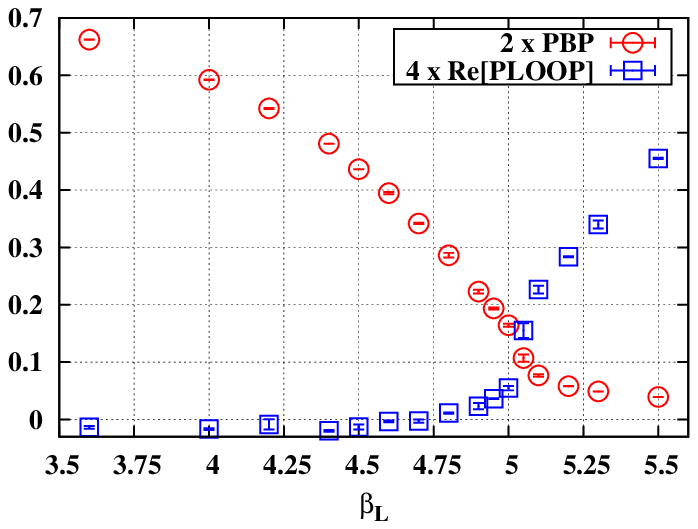}
\includegraphics[width=6.5cm]{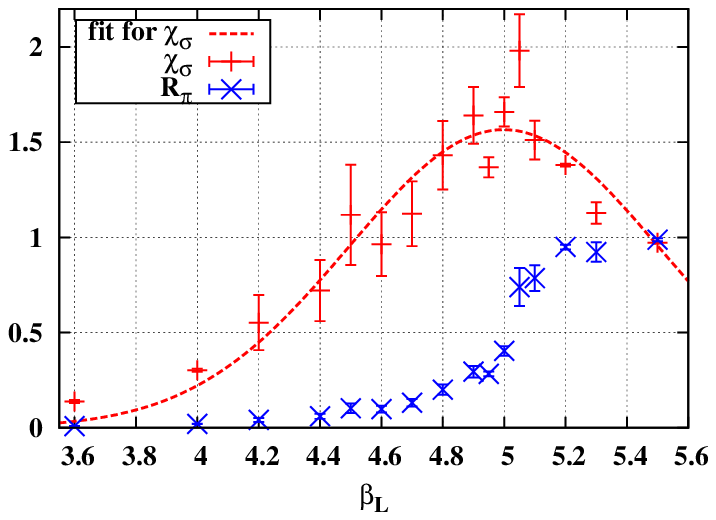}

\includegraphics[width=6.5cm]{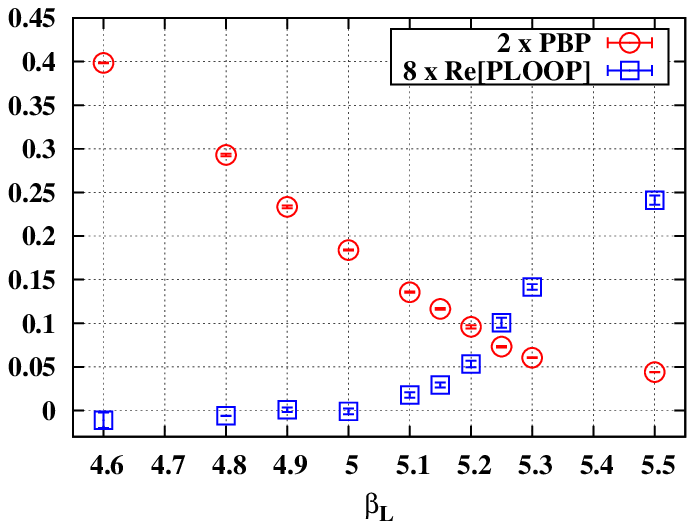}
\includegraphics[width=6.5cm]{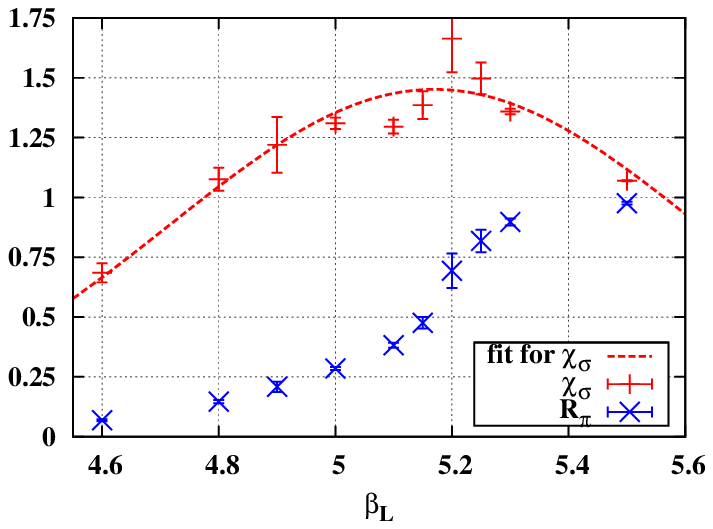}

\includegraphics[width=6.5cm]{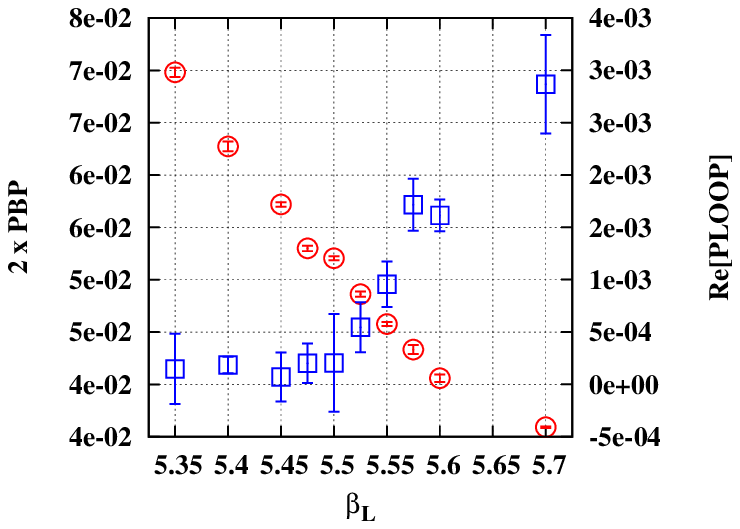}
\includegraphics[width=6.5cm]{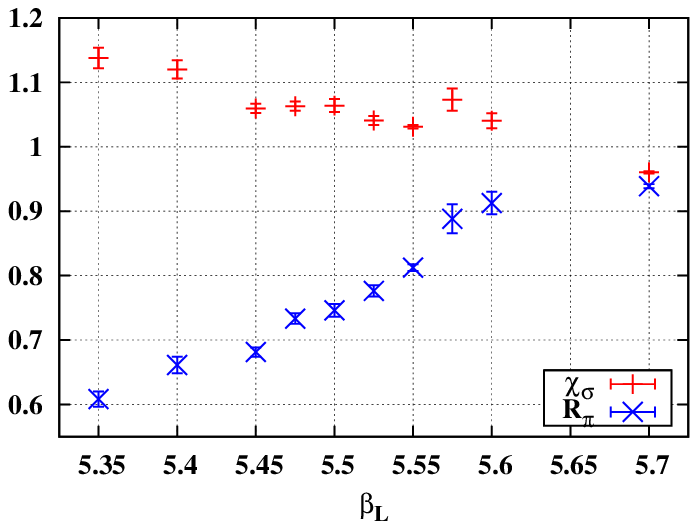}
\caption{The $N_f = 6$ results
for a fixed bare fermion mass $ma = 0.02$.
The first, second, third, and fourth lines
show the results obtained
by using temporal extensions
$N_t = 4,\ 6,\ 8$, and $12$, respectively.
In each line, the left panel shows
the chiral condensate in lattice unit (PBP, red $\bigcirc$)
and the real part of Polyakov loops (Re[PLOOP], blue $\Box$),
and the right panel displays
the chiral susceptibility ($\chi_{\sigma}$, red $+$)
and the chiral susceptibility ratio
($R_{\pi}$, blue $\times$),
as a function of $\betaL$.
For $N_t = 4$, the Gaussian fit for
the chiral susceptibility has been performed in the range
$[4.4,4.9]$ to capture the peak structure.
}\label{Fig:Nf6}
\end{center}
\end{figure*}

We show the $N_f=6$ results
for a fixed bare fermion mass $ma = 0.02$.
In Figs.~\ref{Fig:Nf6},
the chiral condensate
$a^3\langle\bar{\psi}\psi\rangle$ (PBP, red $\bigcirc$)
the real part of Polyakov loop $L$ (Re[PLOOP], blue $\Box$),
the chiral susceptibility ($\chi_{\sigma}$, red $+$),
and the chiral susceptibility ratio
($R_{\pi}$, blue $\times$)
are displayed as a function of a lattice coupling
$\betaL = 10/\gL^2$.
The first, second, third, and fourth lines in the figure
show the results obtained
by using temporal extensions
$N_t = 4,\ 6,\ 8$, and $12$, respectively.
We shall now extract the critical lattice couplings $\betaLC$
associated with the thermal chiral crossover
from these results.

As shown in the left panel of the first line in Fig.~\ref{Fig:Nf6},
the largest decrease of chiral condensates (PBP, red $\bigcirc$)
(as well as a drastic increase of
the real part of Polyakov loops (Re[PLOOP], blue $\Box$))
is found between $\betaL=4.65$ and $4.70$.
Thus, we expect the chiral crossover in this region.
As shown in the right panel of the first line in Fig.~\ref{Fig:Nf6},
the chiral susceptibility $a^2\chi_{\sigma}$ (red $+$)
has a clear peak at $\betaL = 4.65$.
In order to have a practical and
coherent procedure to estimate the maximum,
we have performed Gaussian fits:
The Gaussian fit for the susceptibilities in the range $[4.4,4.9]$
leads to a maximum at a slightly larger $\betaL$ (red dashed line).
Further, the susceptibility ratio $R_{\pi}$ (blue $\times$)
has an inflection point around $\betaL = 4.65 - 4.70$.
For larger $\betaL$, the increasing rate of $R_{\pi}$
significantly reduces, and gets to almost unity.
Thus, all observables consistently indicate
the pseudo-critical coupling to be
$\betaLC = 4.675\pm 0.05$ for $(N_f,N_t) = (6,4)$.
The error is determined
to include the next-to-neighbour data and
the maximum of the Gaussian fit.

The second line in Fig.~\ref{Fig:Nf6}
displays the results for $N_t = 6$.
As shown in the left panel,
the largest decrease of chiral condensates (PBP, red $\bigcirc$)
(as well as a drastic increase of
the real part of Polyakov loops (Re[PLOOP], blue $\Box$))
is found between $\betaL=5.00$ and $5.05$,
and we expect the chiral crossover in this region.
As shown in the right panel,
the chiral susceptibility $a^2\chi_{\sigma}$ (red $+$)
has a peak at $\betaL = 5.05$,
and the Gaussian fit for the susceptibilities
in whole range of $\betaL$
has a maximum at a slightly smaller  $\betaL = 5.0$
(red dashed line).
The susceptibility ratio $R_{\pi}$ (blue $\times$)
has an inflection point around $\betaL = 5.00 - 5.05$,
and then, it goes into the plateau domain.
All observables consistently indicate
the pseudo-critical coupling to be
$\betaLC = 5.025\pm 0.05$ for $(N_f,N_t) = (6,6)$.
The error is determined
to include both $\betaL = 5.0$ and $5.05$ enough.

The third line in Fig.~\ref{Fig:Nf6}
shows the results for $N_t = 8$.
As indicated by the left panel,
the chiral condensates
as well as the Polyakov loops
look smooth at almost everywhere,
and it is difficult to locate the crossover point from them.
As shown in the right panel,
the chiral susceptibility $a^2\chi_{\sigma}$ (red $+$)
has a peak at $\betaL = 5.2$,
and the Gaussian fit for the susceptibilities
in whole range of $\betaL$
has a maximum at a slightly smaller  $\betaL = 5.17$
(red dashed line).
The susceptibility ratio $R_{\pi}$ (blue $\times$)
exhibits the largest variation
between $\betaL = 5.15$ and $5.2$,
after which the increasing rate of $R_{\pi}$
reduces and eventually evolves into almost unity.
From the peak position of $a^2\chi_{\sigma}$,
we estimate the critical coupling to be around
$\betaLC = 5.20\pm 0.05$ for $(N_f,N_t) = (6,8)$.
The error is determined
to include the next neighbour data,
the maximum of the Gaussian fit for $\chi_{\sigma}$,
and the $R_{\pi}$ inflection point.

Finally, we analyse the results for $N_t = 12$,
the largest temporal extension
in our $N_f = 6$ simulations.
The $\betaL$ dependences of chiral condensates
are found to be particularly smooth
for whole range of $\betaL = 4.7 - 5.7$.
Note that in this case,
the aspect ratio ($N_s/N_t$) is only two,
and larger volumes would be required
to reach a comparable clarity in the signal.
As shown in the left panel of final line in Fig.~\ref{Fig:Nf6},
the onset for the Polyakov loop
at $\betaL = 5.525$ is still appreciable (blue $\Box$).
We here notice that the increase of Polyakov loops so far
has been found just before the chiral crossover
in the case of smaller temporal extensions $N_t = 4 - 8$,
though the Polyakov loop itself is not associated
with the chiral dynamics.
Based on such an experience,
we assume that the chiral crossover at $N_t = 12$ is
in the vicinity of the onset of the Polyakov loop,
and carefully investigate the corresponding region
$\betaL = 5.35 - 5.60$.
The chiral condensates do not have any clear signal
(red $\bigcirc$ in the left panel).
As shown in the right panel,
the chiral susceptibility $a^2\chi_{\sigma}$ (red $+$)
has a small peak-like structure at $\betaL = 5.575$,
and a bump-like structure at $\betaL = 5.50$.
The chiral susceptibility ratio $R_\pi$ (blue $\times$)
has the largest increase between $\betaL = 5.55$ and $5.575$,
and tends to be flat in $\betaL \geq 5.575$.
Thus, the critical lattice coupling would be
in the range $5.50 \leq \betaLC \leq 5.575$.
Here, we employ
a conservative estimate $\betaLC = 5.55\pm 0.1$,
which sufficiently covers the whole candidate range.

Our $\betaLC$ collection at $N_f = 6$ is found in
the third line of Table~\ref{Tab:bc}.
As will be shown in the next subsection,
the $N_t$ dependent nature of $\betaLC$ at $N_f = 6$
(a thermal scaling) is associated with the uniqueness of
the physical critical temperature,
indicating the chiral (non-conformal) dynamics at $N_f = 6$.

\subsection{Chiral crossover at $N_f = 8$}
\label{subsec:Nf8}

In our previous paper~\cite{Deuzeman:2008sc},
we have studied the chiral phase transition
at $N_f = 8$ by using two lattice temporal extensions:
$N_t = 6$ and $12$.
One of the main results was
that the chiral phase transition at
$N_f = 8$ still showed
a thermal scaling property,
which indicated the existence of
a typical scale associated with the chiral dynamics
rather than the conformality.
We here add additional data computed at $N_t = 8$,
and confirm the thermal scaling at $N_f = 8$,
for this largish mass.

\begin{figure*}
\begin{center}
\includegraphics[width=6.6cm]{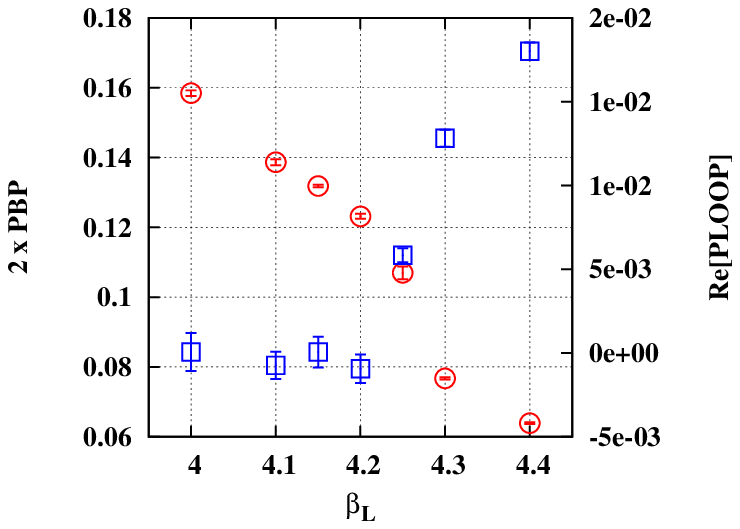}
\includegraphics[width=6.6cm]{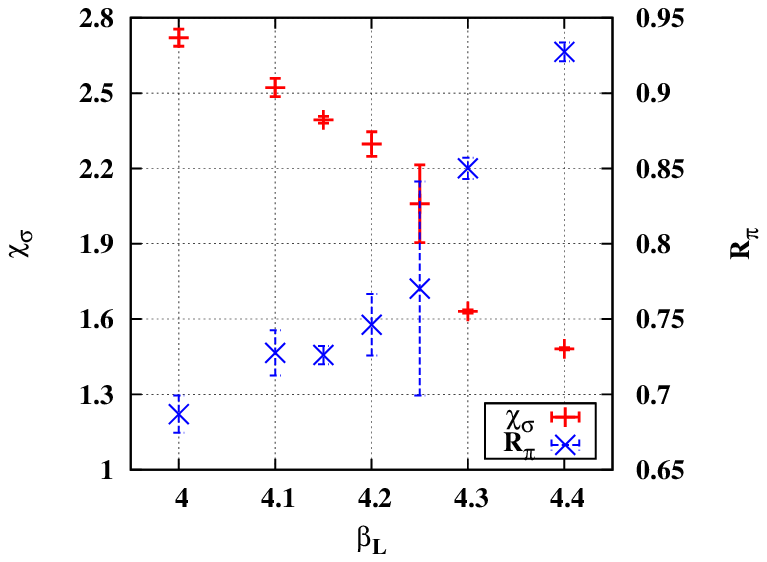}
\caption{The $N_f = 8$ results obtained
by using $24^3\times 8$ lattice volume with $ma = 0.02$:
The chiral condensate in lattice unit (PBP, red $\bigcirc$ in left panel),
the real part of Polyakov loop (Re[PLOOP], blue $\Box$ in left panel),
and the chiral susceptibility ($\chi_{\sigma}$, red $+$ in right panel),
and the chiral susceptibility ratio
($R_{\pi}$, blue $\times$ in right panel)
are shown as a function of $\betaL$.
}\label{Fig:Nf8}
\end{center}
\end{figure*}

The left panel of Fig.~\ref{Fig:Nf8}
shows ensemble averages of chiral condensates
$a^3\langle\bar{\psi}\psi\rangle$ (PBP, red $\bigcirc$),
the real part of Polyakov loop $L$ (Re[PLOOP], blue $\Box$)
as a function of $\betaL$.
We observe the largest decrease of the chiral condensate
between $\betaL = 4.25$ and $4.30$,
while the real part of
the Polyakov loop stars growing around $\betaL = 4.25$.
Although the error is huge,
the chiral susceptibility ratio
$R_{\pi}$ seems to have a larger increase
between $\betaL = 4.25$ and $4.30$.
The large error of $R_{\pi}$ at $\betaL = 4.25$
comes from a very long auto-correlation
in the Monte Carlo trajectories,
which would have required a much larger statistics.
The long correlation hints at a criticality.
All observations consistently indicate
the critical coupling to be $\betaLC = 4.275\pm0.05$.
Combining with $N_t = 6$ and $12$ data~\cite{Deuzeman:2008sc},
we summarise the critical coupling $\betaLC$ at $N_f = 8$
in the final line of Table~\ref{Tab:bc}.

Here we should put some caveats on
the $N_f = 8$ results: 
First, we have not observed a peak-like structure
in the chiral susceptibility $\chi_{\sigma}$.
We should probably study larger spatial volumes, with similar
aspect ratio as those studied in other cases. 
The location
of the pseudo-critical point might change. For the time
being, we rely on the experiences with the other systems,
and infer the pseudo-critical coupling from the other observables.
Second, even in the strong coupling region $\beta < \betaLC$,
$R_{\pi}$ shows relatively large value $\sim 0.7 - 0.8$.
We should go to even stronger coupling and smaller
masses before observing a clear mass gap.
And third, the $N_t$ dependence of $\betaLC$ shows
a large deviation from the two-loop asymptotic scaling law
as will be shown in the next subsection.
This is not surprising, given that the 
couplings explored for $N_f=8$ are larger than in other cases.
This could imply something
which cannot be captured at two-loop,
for example, the pre-conformal dynamics.
Apparently, these caveats call for more
detailed and quantitative lattice studies
with a larger lattice size and a smaller bare fermion mass
before drawing definite conclusions on $N_f=8$.
We note a recent study claiming the conformality
emerges for $N_f=8$ for small enough quark masses
\cite{Schaich:2012fr}.

\subsection{Chiral crossover at $N_f = 4$}
\label{subsec:Nf4}

\begin{figure*}
\begin{center}
\includegraphics[width=6.6cm]{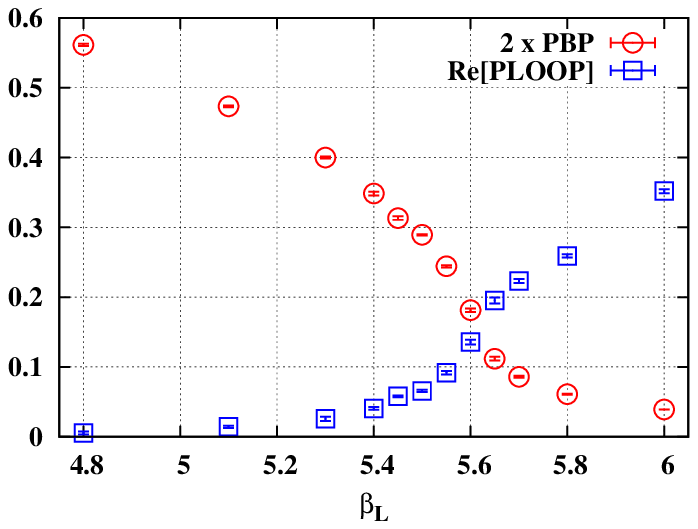}
\includegraphics[width=6.6cm]{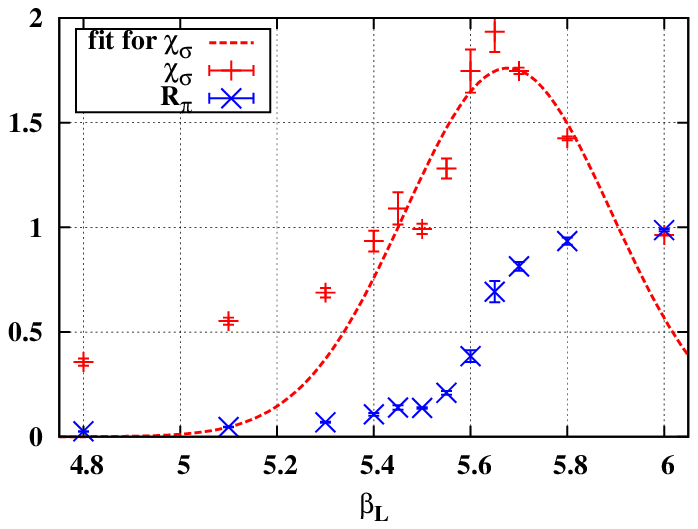}

\includegraphics[width=6.6cm]{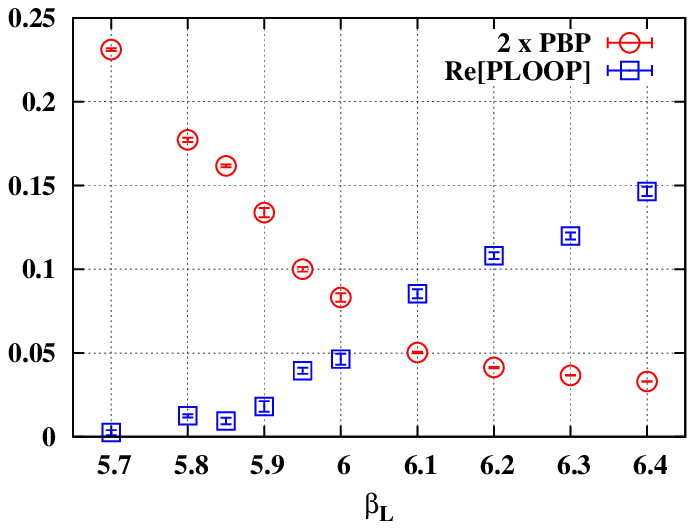}
\includegraphics[width=6.6cm]{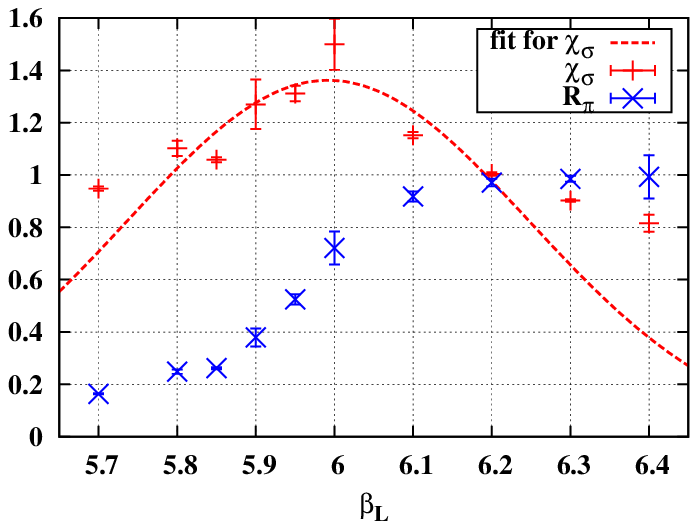}

\includegraphics[width=6.6cm]{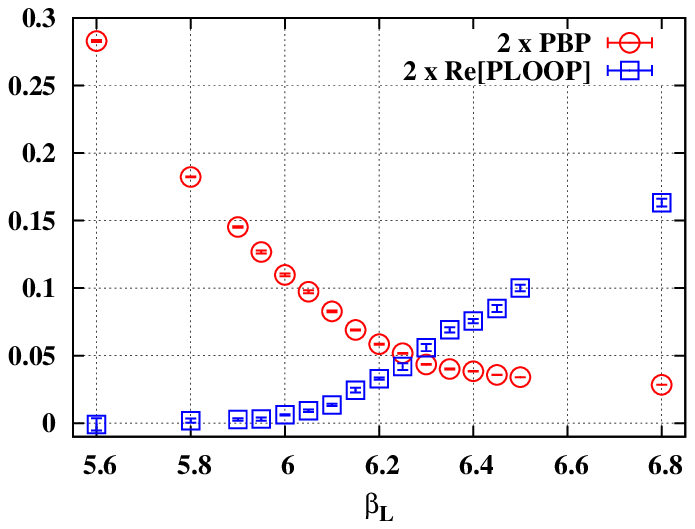}
\includegraphics[width=6.6cm]{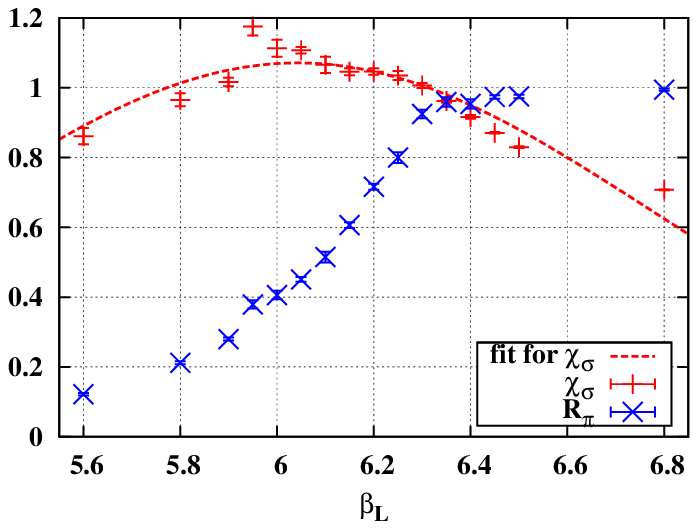}
\caption{The $N_f = 4$ results
for a fixed bare fermion mass $ma = 0.02$.
The first, second, and third lines
show the results obtained
by using temporal extensions
$N_t = 4,\ 6$, and $8$, respectively.
In each line, the left panel shows
the chiral condensate in lattice unit (PBP, red $\bigcirc$)
and the real part of Polyakov loops (Re[PLOOP], blue $\Box$),
and the right panel displays
the chiral susceptibility ($\chi_{\sigma}$, red $+$)
and the chiral susceptibility ratio
($R_{\pi}$, blue $\times$),
as a function of $\betaL$.
{For $N_t = 4$ and $6$, the Gaussian fits for
the chiral susceptibilities have been performed in the range
$[5.4,5.8]$ and $[5.8,6.2]$, respectively.}
}\label{Fig:Nf4}
\end{center}
\end{figure*}

In Figs.~\ref{Fig:Nf4},
we show the chiral condensate
$a^3\langle\bar{\psi}\psi\rangle$ (PBP, red $\bigcirc$),
the real part of Polyakov loop $L$ (PLOOP, blue $\Box$),
the chiral susceptibility ($\chi_{\sigma}$, red $+$),
and the chiral susceptibility ratio ($R_{\pi}$, blue $\times$)
are displayed as a function of a lattice coupling
$\betaL = 10/\gL^2$.
The first, second, and third lines in the figure
show the results obtained
by using temporal extensions
$N_t = 4,\ 6$, and $8$, respectively.

As shown in the left panel of the first line in Fig.~\ref{Fig:Nf4},
the largest decrease of chiral condensates (PBP, red $\bigcirc$)
(as well as a drastic increase of
the real part of Polyakov loops (Re[PLOOP], blue $\Box$))
is found between $\betaL=5.60$ and $5.65$,
and we expect the chiral crossover in this region.
As shown in the right panel of the first line in Fig.~\ref{Fig:Nf4},
the chiral susceptibility $a^2\chi_{\sigma}$ (red $+$)
gets to a maximum at $\betaL = 5.65$.
The Gaussian fit for the susceptibilities in the range $[5.4,5.8]$
leads to a maximum at a slightly larger $\betaL$ (red dashed line).
Further, the susceptibility ratio $R_{\pi}$ (blue $\times$)
has an inflection point around $\betaL = 5.60 - 5.65$.
For larger $\betaL$, the increasing rate of $R_{\pi}$
significantly reduces, and eventually
evolves into almost unity.
Thus, all observables consistently indicate
the pseudo-critical coupling to be
$\betaLC = 5.65\pm 0.05$ for $(N_f,N_t) = (4,4)$.
The error is determined
to include the next-to-neighbour data and
the maximum of the Gaussian fit.

The second line in Fig.~\ref{Fig:Nf4}
displays the results for $N_t = 6$.
As shown in the left panel,
the chiral condensates (PBP, red $\bigcirc$)
are found to be smooth,
and it is difficult to locate the chiral crossover.
The real part of Polyakov loops (Re[PLOOP], blue $\Box$)
starts increasing around $\betaL = 5.95$.
Based on our previous experiences,
the chiral crossover could be around this region.
As shown in the right panel,
the chiral susceptibility $a^2\chi_{\sigma}$ (red $+$)
has a maximum at $\betaL = 6.00$,
and the Gaussian fit for the susceptibilities
in the range of $[5.8,6.2]$
has a maximum at $\betaL = 6.00$ (red dashed line).
From the maximum position of the chiral susceptibilities,
we estimate the pseudo-critical coupling to be
$\betaLC = 6.00\pm 0.05$ for $(N_f,N_t) = (4,6)$.
The error is determined
to include the next-to-neighbour data. 
The susceptibility ratio $R_{\pi}$ (blue $\times$)
has a significant increase in $\betaL > 5.9$,
and then, it flattens at $\betaL = 6.1$.
This behaviour would be consistent
to the above estimate $\betaLC = 6.00$.

The third line in Fig.~\ref{Fig:Nf4}
shows the results for $N_t = 8$.
As indicated by the left panel,
the chiral condensates
as well as the Polyakov loops
look smooth at almost everywhere,
and it is difficult to locate the crossover point from them.
As shown in the right panel,
the chiral susceptibility $a^2\chi_{\sigma}$ (red $+$)
has a peak at $\betaL = 5.95$.
Indeed the susceptibility ratio $R_{\pi}$ (blue $\times$)
also shows a bump structure around $\betaL = 5.95$,
and implies some kinds of an instability of the system.
However, the value of $R_{\pi}$ at $\betaL = 5.95$
turns out to be at most $0.4$,
which indicates a large remaining of the chiral symmetry breaking.
In turn, $R_{\pi}$ keeps increasing
till it approximately reaches to unity at $\betaL = 6.30$,
and thus, the first peak at $\betaL = 5.95$
would not well capture the position of the chiral crossover.
We here postpone the precise determination of the chiral crossover,
and just provide a rough estimate of
the pseudo-critical coupling:
The susceptibility ratio $R_{\pi}$
has a large increasing rate in the range $[6.0,6.3]$.
We adopt the intermediate value as the pseudo-critical coupling
with the error covering whole range of $[6.0,6.3]$,
$\betaLC = 6.15 \pm 0.15$.
This also includes
the maximum of the Gaussian fit
for the chiral susceptibility $\betaL = 6.04$.

The second line of Table~\ref{Tab:bc} provides
a summary of $\betaLC$ for $N_f = 4$.

\subsection{Deconfinement  at $N_f = 0$}
\label{subsec:Nf0}
In this subsection, we estimate
the critical lattice coupling $\betaLC$
for deconfinement in the quenched ($N_f = 0$) system.
We note that both deconfinement and chiral transitions
are associated with the thermal phase transition
from the hadronic phase to the non-Abelian plasma phase
with a drastic increase of the pressure (degrees of freedom).
Then, our interest is the probe of the system with various $N_f$
in light of such a thermal phase transition.
In this sense, we regard the deconfinement crossover
at $N_f = 0$ as a {continuation}
to the chiral crossover at finite $N_f$.
In our setup, this connection is made explicit by the
fact that we are realising a quenched system by use of a large
mass in the four flavour system. It should be noted, anyway,
that our result for the (pre-)conformal dynamics
does not crucially depend on the quench{ed} data.

\begin{figure*}
\begin{center}
\includegraphics[width=6.6cm]{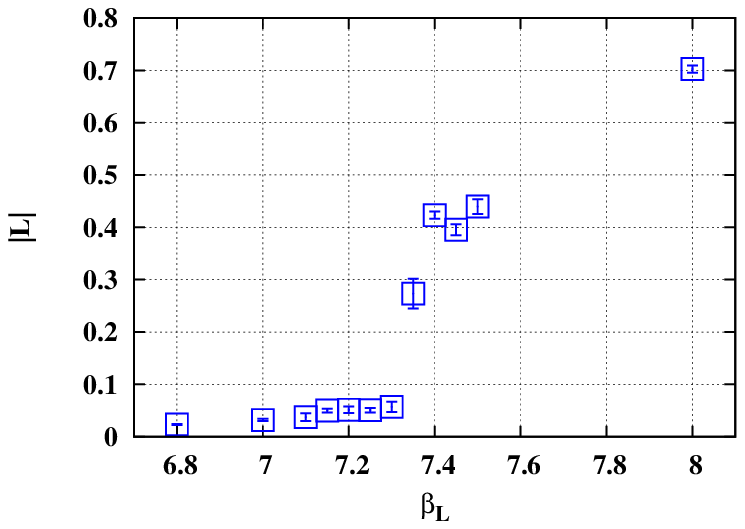}
\includegraphics[width=6.6cm]{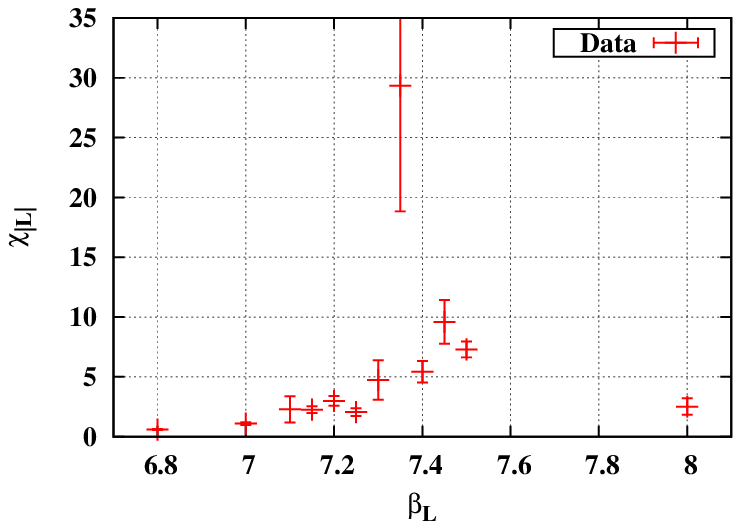}

\includegraphics[width=6.6cm]{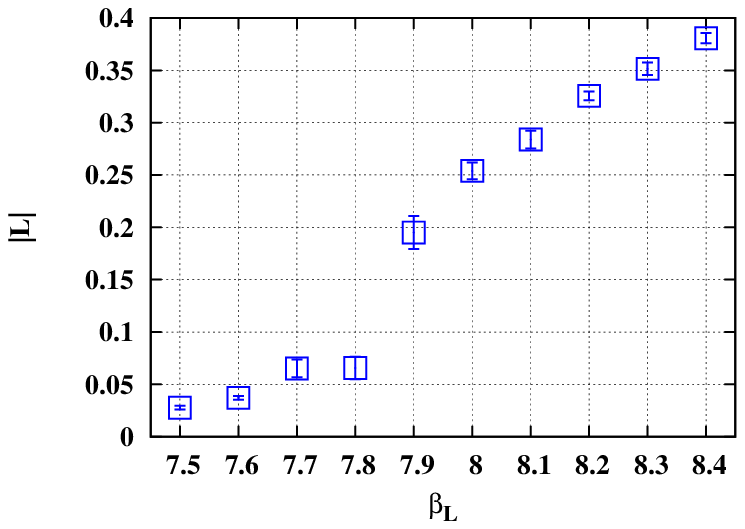}
\includegraphics[width=6.6cm]{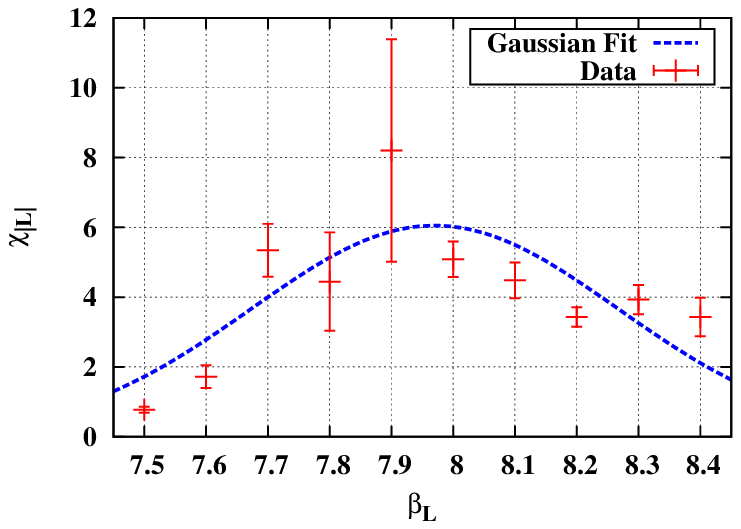}

\includegraphics[width=6.6cm]{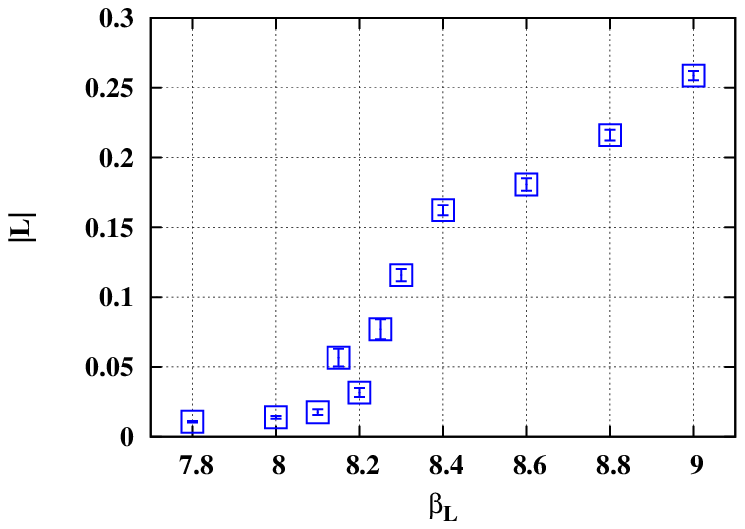}
\includegraphics[width=6.6cm]{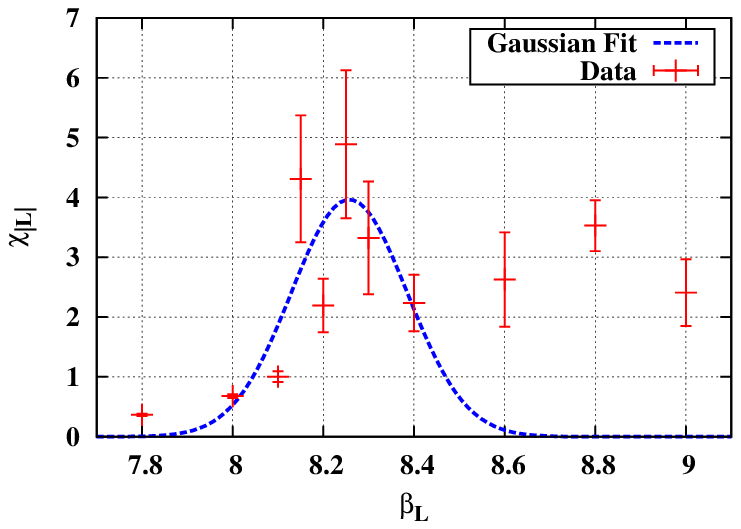}
\caption{The quenched results.
The first, second, and third lines
show the results obtained
by using temporal extensions
$N_t = 4,\ 6$, and $8$, respectively.
In each line, the left panel shows
the absolute value of Polyakov loops ($|L|$, $\Box$),
and the right panel displays
the susceptibility calculated from the variance of $|L|$
($\chi_{|L|}$, symbol $+$).
The Gaussian fits are particularly
bad here, however the identification of the maximum is clear.
For $N_t = 8$, the fit has been performed excluding the data
in $\betaL > 8.8$, because as shown in the left panel,
the drastic increase of $|L|$
is found {for a} much smaller $\betaL$.
}\label{Fig:Nf0}
\end{center}
\end{figure*}

In Figs.~\ref{Fig:Nf0},
the thermalized ensemble averages of
the absolute of Polyakov loop ($|L|$, blue $\Box$),
its susceptibility ($\chi_{|L|}$, red $+$),
are displayed as a function of a lattice coupling
$\betaL = 10/\gL^2$.
The first, second, and third lines in the figure
show the results obtained
by using temporal extensions
$N_t = 4,\ 6$, and $8$, respectively.

As shown in the left panel of the first line in Fig.~\ref{Fig:Nf0},
the largest increase of 
the absolute value of Polyakov loops ($|L|$, blue $\Box$)
is found between $\betaL=7.30$ and $7.35$,
and we expect the deconfinement crossover in this region.
As shown in the right panel,
the susceptibility for $|L|$ (red $+$)
has a clear peak at $\betaL = 7.35$,
hence we estimate
the pseudo-critical coupling to be
$\betaLC = 7.35\pm 0.05$ for $(N_f,N_t) = (0,4)$.
The error is determined
to include the next-to-neighbour data.

The second line in Fig.~\ref{Fig:Nf0}
displays the results for $N_t = 6$.
As shown in the left panel,
the largest increase of 
the absolute value of Polyakov loops ($|L|$, blue $\Box$)
is found between $\betaL=7.80$ and $7.90$,
and we expect the deconfinement crossover in this region.
{The maximum of 
the susceptibility evaluated from $|L|$ (red $+$)
is observed at $N_f = 7.9$. The large error indicates
the long correlation time of the Monte Carlo trajectories.
The Gaussian fit for the susceptibility  
has a maximum at $\betaL = 7.97$.}
From this, we estimate the pseudo-critical coupling to be
$\betaLC = 7.97\pm 0.07$ for $(N_f,N_t) = (0,6)$.
The error is determined to include
the next-to-next-to-neighbour data from the maximum point.

The third line in Fig.~\ref{Fig:Nf0}
represents the results for $N_t = 8$.
As shown in the left panel,
the absolute value of Polyakov loop ($|L|$, blue $\Box$)
starts increasing around $\betaL = 8.15$, and
we expect the deconfinement crossover in this region.
{The maximum of 
the susceptibility evaluated from $|L|$ (red $+$)
is observed at $N_f = 8.25$.
Again, we find the large error in the vicinity
of the maximum, indicating
the long correlation time of the Monte Carlo trajectories.
The Gaussian fit for the susceptibility in the range $[7.8,8.6]$
has a maximum at $\betaL = 8.26$.
Note that the fit range sufficiently covers whole region of
the drastic increases of $|L|$ shown in the left panel.}
We adopt the maximum of the Gaussian fit
as a critical coupling:
$\betaLC = 8.26\pm 0.06$ for $(N_f,N_t) = (0,6)$.
The error is determined to include
the next-to-next-to-neighbour data from the maximum point.

The first line of Table~\ref{Tab:bc} provides
a summary of $\betaLC$ for $N_f = 0$.
The $N_t$ dependent nature of $\betaLC$
reflects the thermal nature of the crossover.

\section{Asymptotic scaling analyses for chiral phase transition}
\label{sec:AS}
In this section,
we investigate the asymptotic scaling of
the pseudo-critical temperatures $T_c$
\begin{align}
&T_c\equiv \frac{1}{a(\betaLC)\cdot N_t}\ .\label{eq:Tc}
\end{align}
where $\betaLC$ have been computed  in the previous section,
and discuss the connection to the continuum physics.
In the first subsection \ref{subsec:TcL},
we introduce the normalised critical temperature
$T_c/\Lambda_{\mathrm{L/E}}$ 
(see e.g. \cite{Gupta:2000hr})
where $\Lambda_{\mathrm{L}}$ ($\Lambda_{\mathrm{E}}$)
represents the lattice (E-scheme) Lambda-parameter
defined in the two-loop perturbation theory
with or without a renormalisation group inspired
improvement~\cite{CAllton}.
Then in the subsections \ref{subsec:TcL_Nt},
the asymptotic scaling will be assessed
by studying the $N_t$ (in)dependence
of $T_c/\Lambda_{\mathrm{L/E}}$ for each $N_f$.

\subsection{Normalised critical temperature
$T_c/\Lambda_{\mathrm{L/E}}$}\label{subsec:TcL}
We  consider the two-loop beta-function
\begin{align}
&\beta(g)
=-(b_0 {g}^3 + b_1 {g}^5)\ ,\label{eq:beta_func}\\
&b_0
=
\frac{1}{(4\pi)^2}
\Biggl(
\frac{11C_2[G]}{3}-\frac{4T[F]N_f}{3}
\Biggr)\ ,\label{eq:b0}\\
&b_1
=
\frac{1}{(4\pi)^4}
\Biggl(
\frac{34(C_2[G])^2}{3}
-\biggl(\frac{20C_2[G]}{3}+4C_2[F]\biggr)T[F]N_f
\Biggr)\ ,\label{eq:b1}
\end{align}
with $(C_2[G],\,C_2[F],\,T[F])=(N_c,\,(N_c^2-1)/(2N_c),\, 1/2)$.
The coupling $g$ can be either 
the lattice bare coupling
$\gL = \sqrt{10/\betaL}$ or the E-scheme renormalised coupling
$\gE = \sqrt{3(1-\langle P \rangle(\gL))}$,
where $\langle P\rangle(\gL)$ is the zero temperature
plaquette value. 
If the one-loop perturbation theory exactly holds,
the E-scheme coincides the lattice scheme.

Integrating Eq.~(\ref{eq:beta_func}),
we obtain the well-known two-loop asymptotic scaling relation,
\begin{align}
R(\gLE)\equiv
a(\gLE)\Lambda_{\mathrm{L/E}}
= \bigl(b_0\gLE^2\bigr)^{-b_1/(2b_0^2)}
\exp\biggl[
\frac{-1}{2b_0\gLE}
\biggr]
\ ,\label{eq:RL}
\end{align}
where  $\Lambda_{\mathrm{L}}$ ($\Lambda_{\mathrm{E}}$)
is the Lattice (E-scheme) Lambda-parameter. 

To take into account  higher order corrections,
we have also considered the renormalisation group inspired
improvement~\cite{CAllton}
\begin{align}
R^{\mathrm{imp}}(\betaLE)=
\LEI~a(\betaLE)
\equiv
\frac{R(\betaLE)}{1+h}
\times
\Biggl[
1 + h\
\frac{R^2(\betaLE)}{R^2(\beta_0)}
\Biggr]\ ,
\label{eq:RL_imp}
\end{align}
where $\betaLE = 10/(\gLE)^2$.
The coupling $\beta_0$  can be arbitrarily set 
and the parameter $h$ is adjusted
so as to minimise the scaling violation.
Note that $h = 0$ reproduces the standard asymptotic scaling law
Eq.~(\ref{eq:RL}).

\begin{figure}
\begin{center}
\includegraphics[width=6.6cm,height=6.0cm]{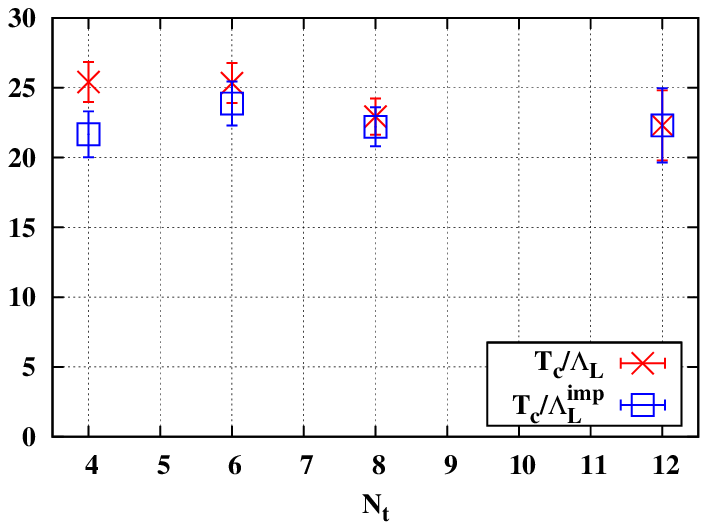}
\includegraphics[width=6.6cm,height=6.0cm]{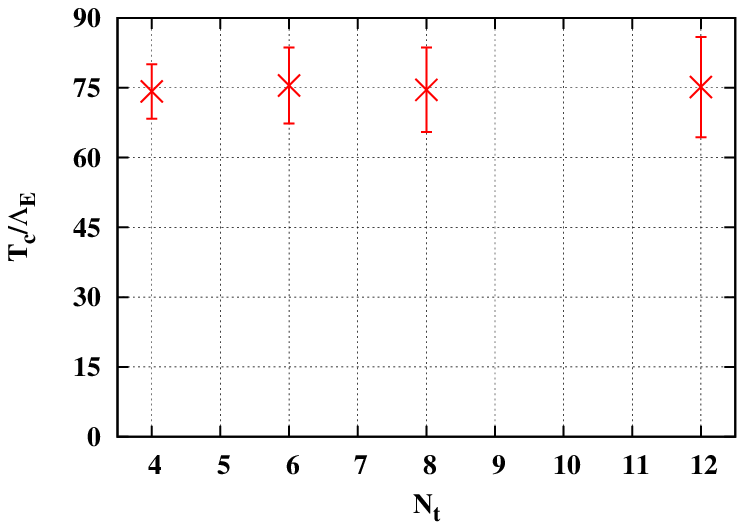}
\caption{Scaling at $N_f = 6$
from the $N_t$ dependence of the normalised
critical temperature.
Left: The bare lattice scheme results.
The red symbol $\times$ shows $T_c/\Lambda_{\mathrm{L}}$,
and the blue $\Box$ symbols represent $T_c/\LI$
obtained by using the parameters
$h = 0.03$ and $\beta_0 = \betaLC(N_f = 6, N_t = 12) = 5.55$
in Eq.~(\protect\ref{eq:RL_imp}).
Right: The E-scheme results
$T_c/\Lambda_{\mathrm{E}}$.}
\label{Fig:TcL_Nf6}
\end{center}
\end{figure}

\begin{figure}
\begin{center}
\includegraphics[width=6.6cm,height=6.0cm]{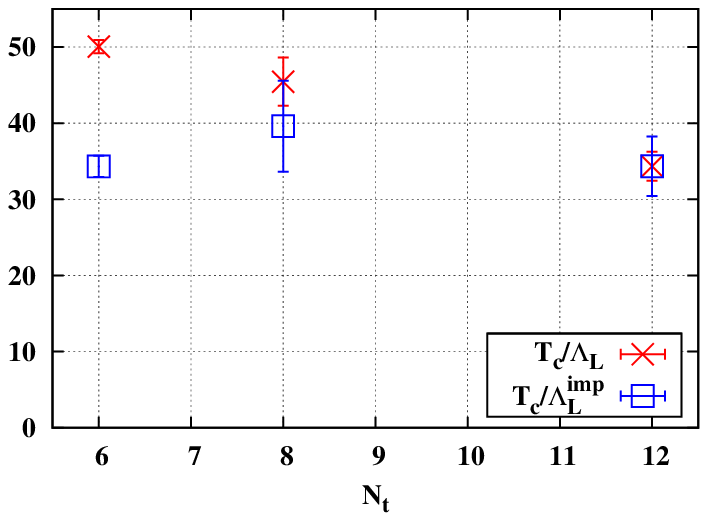}
\includegraphics[width=6.6cm,height=6.0cm]{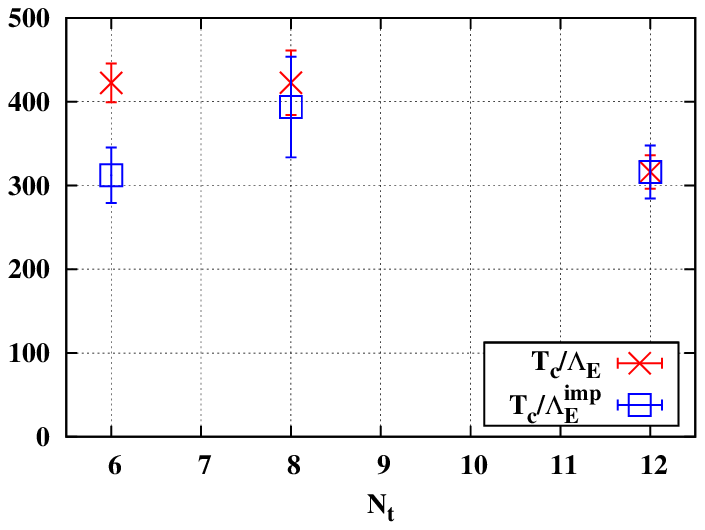}
\caption{Scaling at $N_f = 8$
from the $N_t$ dependence of the normalised
critical temperature.
Left: The bare lattice scheme results.
The red symbol $\times$ shows $T_c/\Lambda_{\mathrm{L}}$,
and the blue $\Box$ symbols represent $T_c/\LI$
obtained by using the parameters
$h = 1.08$ and $\beta_0 = \betaLC(N_f = 8, N_t = 12) = 4.34$
in Eq.~(\protect\ref{eq:RL_imp}).
Right: The E-scheme results.
The red symbol $\times$ shows $T_c/\Lambda_{\mathrm{L}}$,
and the blue $\Box$ symbols represent $T_c/\EI$
obtained by using the parameters
$h = 0.4$ and $\beta_0 = \betaEC(N_f = 8, N_t = 12) \simeq 5.94$
in Eq.~(\protect\ref{eq:RL_imp}).}
\label{Fig:TcL_Nf8}
\end{center}
\end{figure}

The asymptotic scaling as described above
is valid in the massless limit.
In the following,
we will use it to analyse results obtained
at finite bare fermion mass $ma = 0.02$
by assuming that the shift of the 
(pseudo) critical coupling induced by a non-zero mass
is smaller than other errors.
This assumption should ultimately
be tested in future studies
by performing simulations with different masses and
extrapolating to the chiral limit.

We now substitute $\betaLEC$
into the temperature definition Eq.~(\ref{eq:T}),
and insert the scale $\Lambda_{\mathrm{L/E}}$:
\begin{align}
\frac{1}{N_t}=\frac{T_c}{\Lambda_{\mathrm{L/E}}}
\times \Bigl(\Lambda_{\mathrm{L/E}}~a(\betaLEC)\Bigr)
\ .\label{eq:T_Lam}
\end{align}
The left-hand side is a given number,
and $\Lambda_{\mathrm{L/E}}~a(\betaLEC)$
in the right-hand side
is evaluated by using Eq.~(\ref{eq:RL})
and our critical couplings $\betaLEC$.
Thus, Eq.~(\ref{eq:T_Lam}) allows us
to convert the critical couplings
into the (normalised) critical temperature
$T_c/\Lambda_{\mathrm{L/E}}$.
When we adopt the improvement Eq.~(\ref{eq:RL_imp}),
$T_c/\Lambda_{\mathrm{L/E}}$ is upgraded into
$T_c/\LEI$.

\subsection{Results for $T_c/\Lambda_{\mathrm{L/E}}$ and $T_c/\LEI$
}\label{subsec:TcL_Nt}

In this section we consider 
\begin{align}
\frac{T_c}{\Lambda_{\mathrm{L/E}}}
=\frac{R(\gLE)}{N_t}
= \bigl(b_0\gLE^2\bigr)^{-b_1/(2b_0^2)}
\exp\biggl[
\frac{-1}{2b_0}
\biggr]
\ ,\label{eq:TcL}
\end{align}
where $\gLE$ denotes either the bare lattice coupling
or the coupling defined in the E scheme. 
In addition, we consider the renormalisation group inspired 
definition,
\begin{align}
\frac{T_c}{\LEI}
= \frac{R^{\mathrm{imp}}(\gLE)}{N_t}
\ ,\label{eq:TcL_imp}
\end{align}
where $R^{\mathrm{imp}}$ is given by Eq.~(\ref{eq:RL_imp}).
The numerical results for
$T_c/\Lambda_{\mathrm{L/E}}$ and $T_c/\LEI$ 
are collected in   
Table \ref{tab:TcL} and Table \ref{tab:TcLE}.

\begin{table*}
\caption{
Summary of
$T_c/\Lambda_\mathrm{L}$ and
$T_c/\LI$ for various $(N_f,N_t)$.
The first (second) line at fixed $(N_f,N_t)$
shows the value of $T_c/\Lambda_\mathrm{L}$ ($T_c/\LI$),
and the last two columns provide
the parameter $h$ and $\beta_0$ appeared
in the improved asymptotic scaling
Eq.~(\protect\ref{eq:RL_imp}).
}\label{tab:TcL}
\begin{center}
\begin{tabular}{c|cccc|cc}
\hline\hline
$N_f\backslash N_t$ &
$4$&
$6$&
$8$&
$12$&
$h$&
$\beta_0$\\
\hline
$0$ &
$18.11\pm 0.65$&
$18.21\pm 0.91$&
$16.56\pm 0.71$&
$-$&
$-$&
$-$\\
\quad &
$16.29\pm 0.75$&
$17.81\pm 1.02$&
$16.56\pm 0.78$&
$-$&
$0.05$&
$8.26$\\
\hline
$4$&
$21.99\pm 1.04$&
$19.98\pm 0.95$&
$17.12\pm 2.43$& 
$-$&
$-$&
$-$\\
\quad &
$16.56\pm 1.44$&
$18.67\pm 1.38$&
$17.12\pm 3.41$&
$-$&
$0.30$&
$6.15$\\
\hline
$6$ &
$25.41\pm 1.43$&
$25.33\pm 1.43$&
$22.94\pm 1.29$&
$22.30\pm 2.52$&
$-$&
$-$\\
\quad &
$21.66\pm 1.64$&
$23.87\pm 1.58$&
$22.21\pm 1.40$&
$22.30\pm 2.66$&
$0.03$&
$5.55$\\
\hline
$8$ &
$-$&
$50.05\pm 0.87$&
$47.06\pm 3.28$&
$34.34\pm 1.91$&
$-$&
$-$\\
\quad &
$-$&
$34.32\pm 1.40$&
$42.67\pm 6.33$&
$34.34\pm 3.90$&
$1.08$&
$4.34$\\
\hline\hline
\end{tabular}
\end{center}
\end{table*}

\subsubsection{$N_f=6$}
The left panel of Fig.
~\ref{Fig:TcL_Nf6}
shows $T_c/\Lambda_{\mathrm{L}}$ 
as a function of  $N_t$ for $N_f = 6$, without
( red ($\times$) symbols) and with 
(blue ($\Box$) symbols) improvement. 
The improvement parameter $h=0.03$ is adjusted to minimise 
the $N_t$ dependence. We have checked that
the results are stable against small variation of $h$. Moreover
$\beta_0$ is  adjusted to match the results at 
$\beta_0  = \betaLC (N_t = 12) = 5.55$.
Similarly, the right panel of Fig. ~\ref{Fig:TcL_Nf6} shows 
$T_c/\Lambda_{\mathrm{E}}$ , which is nearly
constant with $N_t$. 
The overall behaviour suggests that the residual scaling
violations at $N_t =12$ are small. 

\subsubsection{$N_f=8$}
In the left panel of Fig. \ref{Fig:TcL_Nf8},
we show the $N_t$ dependence of
the normalised critical temperature in the lattice scheme, 
without ( red ($\times$) symbols) and with 
(blue ($\Box$) symbols) improvement. Again we 
 adjust the improvement parameter $h$ so to minimise 
the $N_t$ dependence, and we find that a larger $h \simeq 1.08$
is needed. This is consistent with the larger scaling violation
observed between $N_t=6$ and $N_t=12$.  Similar observations
can be made on the E-scheme,
where the scaling violations, as seen in the $N_t$
dependence,  appear to be larger than those
in $N_f=6$ case. Introducing the improvement in the
E-scheme  again requires a largish 
$h = 0.4$ ($\beta_0 = \betaEC(N_f = 8, N_t = 12) = 5.94$).

In short summary,
the system with $N_f = 8$ shows much larger and 
less controllable deviations
from the two-loop asymptotic scaling, than the ones observed
for $N_f = 6$ (and for $N_f=4$ and $N_f=0$, to be  discussed next). 
This might be natural, in view of the largish
values of the coupling involved. 
These observations confirm that in this
case the $\beta$ function
at two loops cannot offer a quantitative guidance to
strongly coupled pre-conformal dynamics.

\subsubsection{$N_f=0$ and $N_f=4$}
For $N_f=0$,
we find about $10$ percent variation of
$T_c/\Lambda_\mathrm{L}$
in the whole range $N_t \in [4,8]$
(In this case  $T_c$ represents
the critical temperature associated
with the deconfinement transition
rather than the chiral transition).
The $N_t$ dependence can be reduced to
less than $10$ percent
by using the improved  scaling
with a small $h=0.05$.
Turning to $N_f=4$ 
we find about $20$ percent variations
of $T_c/\Lambda_\mathrm{L}$
between the $N_t = 4$ and the $N_t = 8$ results.
The improved asymptotic scaling Eq.~(\ref{eq:RL_imp}) works well,
and the variation reduces into $10$ percent level
in whole range of $N_t = 4 - 8$.
\begin{table*}
\caption{
Summary of $T_c/\Lambda_\mathrm{E}$
and $T_c/\LEI$ for $N_f = 6$ and $N_f = 8$.
The first (second) line at fixed $(N_f,N_t)$
shows the value of $T_c/\Lambda_\mathrm{E}$ ($T_c/\LEI$),
and the last two columns give
the parameter $h$ and $\beta_0$ appeared
in the improved asymptotic scaling Eq.~(\protect\ref{eq:RL_imp}).
For $N_f = 6$, the improvement was not necessary.
}\label{tab:TcLE}
\begin{center}
\begin{tabular}{c|cccc|cc}
\hline\hline
$N_f\backslash N_t$ &
$4$&
$6$&
$8$&
$12$&
$h$&
$\beta_0$\\
\hline
$6$ &
$74.22\pm 5.86$&
$75.47\pm 8.17$&
$74.56\pm 9.08$&
$75.13\pm 10.76$&
$-$&
$-$\\
\hline
$8$ &
$-$&
$422.54\pm 23.06$&
$422.61\pm 38.59$&
$316.03\pm 20.06$&
$-$&
$-$\\
\quad &
$-$&
$312.16\pm 33.13$&
$393.58\pm 60.01$&
$316.03\pm 31.52$&
$0.40$&
$4.34$\\
\hline\hline
\end{tabular}
\end{center}
\end{table*}
\subsubsection{Scale separation?}
In summary, $T_c/\Lambda$ computed using different schemes
($\Lambda = \Lambda_{\mathrm{L}}$ or $\Lambda_{\mathrm{E}}$)
consistently shows an increase with $N_f$, confirming 
and extending the
findings of our early work \cite{Miura:2011mc}.
As discussed in  \cite{Miura:2011mc} this indicates that 
$\Lambda_{\mathrm{L/E}}$  vanishes faster than $T_c$
upon approaching the critical number of flavour. Within the various
uncertainties discussed here, this can be taken as a qualitative
indication of  scale separation close to the critical
number of flavors.

\section{Onset of the conformal window}\label{sec:discuss}

In this section,
we study the emergence of the
conformal phase with increasing $N_f$.
In the first Subsection \ref{subsec:MY},
we consider the phase diagram 
in the space spanned by the bare coupling $\gL$
and the number of flavor $N_f$.  
We consider the (pseudo)critical thermal lines which connect
the lattice (pseudo)critical couplings for a  fixed $N_t$ . 
We argue that
the critical number of flavor $N_f^*$  can be identified
with the crossing point of the pseudo-critical thermal
lines obtained for various $N_t$'s. 

In the second Subsection \ref{subsec:gTC},
we introduce the thermal critical coupling $\gTC(N_f)$
as a typical interaction strength at
the scale of the critical temperature $T_c(N_f)$.
Since $T_c$ approaches zero when the number of flavor
approaches the lower edge of conformal window $N_f^*$,
$\gTC (N_f = N_f^*)$ should be equal to
the zero temperature critical coupling $g^c$
(to be specified and estimated in the following).
The relation $\gTC(N_f^*) = g^c$ thus defines implicitly
the critical number of flavor $N_f^*$. 

In the final Subsection \ref{subsec:TcM},
we develop an improved version
of the approach used in our early paper \cite{Miura:2011mc}. 
We introduce a UV $N_f$ independent reference scale,
and compute the critical temperature for each $N_f$
in units of this scale. The infra-red dynamics affecting 
the critical temperature $T_c$
is then clearly exposed, and 
$N_f^*$ can be estimated from the vanishing of $T_c$. 

Before turning to the details, 
we summarise our results for $N_f^*$:
\begin{align}
N_f^* \sim 
\begin{cases}
11.1\pm 1.6\quad &
(\text{from the lattice thermal lines} )\ ,\\
12.5\pm 1.6\quad &
(\text{from the strength of the coupling at $T_c$})\ ,\\
10.4\pm 1.2\quad &
(\text{from the vanishing of $T_c$})\ .
\end{cases}
\label{eq:Nf_IRFP}
\end{align}

\subsection{The critical number of flavor from the
lattice thermal lines}\label{subsec:MY}

\begin{figure}
\begin{center}
\includegraphics[width=9.5cm]{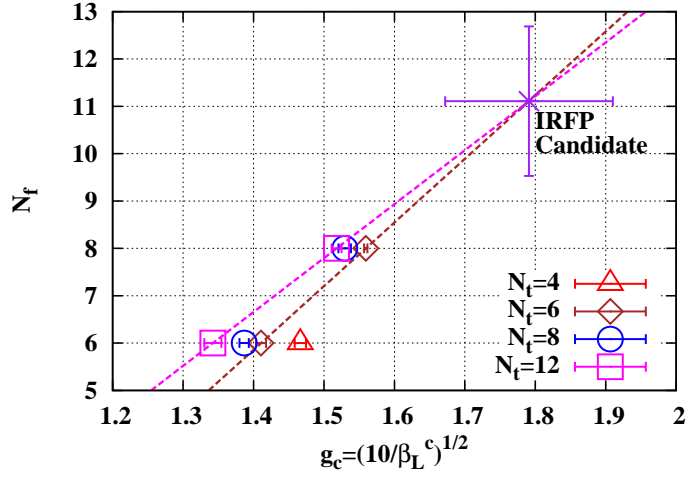}
\caption{(Pseudo) critical values of the lattice coupling
$\gLC=\sqrt{10/\betaLC}$ for theories with $N_f=0,~4,~6,~8$ 
and for several values of $N_t$
in the Miransky-Yamawaki phase diagram.
We have picked up $\gLC$ at $N_f = 6$ and $8$,
and considered  ``constant $N_t$'' lines
with $N_t = 6,\ 12$.
{If the system is still described by
one parameter beta-function in this range of coupling,
the IRFP could be located at the intersection of the
fixed $N_t$ lines -- or equivalently, in the region where
the step-scaling function vanishes. To demonstrate the procedure
--as a preliminary example -- 
we have considered the intersection
of the $N_t = 12$ and $N_f = 6$ lines}.}
\label{Fig:MY}
\end{center}
\end{figure}

We plot the lattice critical couplings
$\gLC(N_f,N_t) = \sqrt{10/\betaLC(N_f,N_t)}$
(Table \ref{Tab:bc})
in the space spanned by the bare coupling $\gL$
and the number of flavor $N_f$, and 
we consider the lines which connect
$\gLC$ with $N_t$ fixed: $\gLC(N_f)|_{N_t=\mathrm{fix}}$
(see Fig.~\ref{Fig:MY}).
These pseudo-critical thermal lines 
separate a phase where chiral symmetry
(approximately) holds
from a phase where chiral symmetry is spontaneously broken
\footnote{
It would be of interest to study the interrelation of such lines
with the zero temperature first order transition line
observed in the conformal window
\cite{Deuzeman:2012ee,Deuzeman:2011pa,Deuzeman:2012pv,Deuzeman:2009mh,Schaich:2012fr,Cheng:2011ic,Hasenfratz:2010fi,Hasenfratz:2011xn,Damgaard:1997ut,deForcrand}.}.
The resultant phase diagram may be seen
as an extension of the well-known
Miransky-Yamawaki phase diagram \cite{Miransky:1997}
to finite temperature.

We here argue that
the critical number of flavor $N_f^*$ can be read off
from the crossing point of thermal lines obtained for different $N_t$.
To see this,
we consider the well-known step-scaling function: 
\begin{equation}
\Delta\betaL^s = \betaL - {\betaL}^{\prime}
\end{equation}
where $\betaL$ and $\betaL^{\prime}$ 
give the same physical scale $\xi$:
\begin{align}
\xi = a(\betaL)\hat{\xi} = a({\betaL}^{\prime})\hat{\xi}^{\prime}
\ .\label{eq:unique_xi}
\end{align}
Here, $\hat{\xi}$ is the dimension-less
lattice correlation length, and 
$\hat{\xi}/ \hat {\xi^{\prime}} = s$.
In our case, $\xi = T_c^{-1}$, $\hat{\xi} = N_t, 
\hat{\xi}^{\prime} = N_t^{\prime}$,
and the above relation Eq.~(\ref{eq:unique_xi}) reads 
\begin{align}
T_c^{-1} = N_t\ a(\betaLC) = N_t^{\prime}\ a({\betaLC}^{\prime})
\ .\label{eq:unique_Tc}
\end{align}

As discussed in the previous study Ref.~\cite{Hasenfratz:2011xn},
$\Delta \betaL^s = 0$ holds at the IRFP
regardless the scale factor $s$.
In principle,
we could then compute the step-scaling function from
our numerical results,
and try to see where it vanishes.
Alternatively,
we can look for the intersection
of pseudo-critical thermal lines:
obviously, $\Delta  \betaL^s = 0$ holds at the intersection point 
regardless the value of the scale factor $s$.

To demonstrate this procedure,
we consider
the pseudo-critical lines obtained for $N_t = 6$ and $N_t=12$
as shown in Fig.~\ref{Fig:MY}.
Note their positive slope:
the lattice critical coupling $\gLC$ is
an increasing function of $N_f$.
This is a consequence
of enhanced fermionic screening for a large number
of flavor, as noted first in Ref.~\cite{Kogut:1985pp}.
Interestingly, the slope decreases with increasing $N_t$,
which allows for a crossing point at a larger $N_f$.
Thus, we estimate the intersection at
$(\gLC, N_f^*) = (1.79\pm 0.12,  11.1\pm 1.6)$.

We underscore at this stage that the above analysis
is merely a qualitative discussion:
the precise shape of the pseudo-critical thermal lines
with fixed $N_t$ is dictated by the
beta-function, which is unknown. Hence, a linear
extrapolation which only uses two values of $N_t$ 
has only the meaning to illustrate
a viable strategy which we plan to further pursue 
in the future. This caveat issued, the agreement of the results
found here with the estimates presented below is
rather gratifying.

\subsection{The critical number of flavor from the 
interaction strength at $T_c$}
\label{subsec:gTC}

In this second subsection,
we will follow 
the approach of a recent paper \cite{Liao:2012tw},
and compute the coupling $\gTC (N_f)$ at the scale of
the critical temperature for each $N_f$.

To obtain the coupling at the scale of the temperature,
we evolve the coupling at the scale of the lattice spacing
$a$ 
up to the temperature inverse scale $N_t a$.
To this end,
we make use of two-loop expressions.
Consider the renormalisation flow:
\begin{align}
&\bar{R}(\gLC,\gLR)
\equiv \frac{M(\gLR)}{a^{-1}(\gLC)}
=\exp\biggl[
\int_{\gLC}^{\gLR}\frac{d\gL}{\beta(\gL)}
\biggr]\nn\\
&\simeq
\Biggl(
\frac{(\gLC)^2}{(\gLC)^2b_1 + b_0}
\frac{(\gLR)^2b_1 + b_0}{(\gLR)^2}
\Biggr)^{-b_1/(2b_0^2)}\nn\\
&\qquad\times
\exp\Biggl[
\frac{1}{2b_0}
\biggl(\frac{1}{(\gLR)^2} - \frac{1}{(\gLC)^2}\biggr)
\Biggr]\ ,\label{eq:RV}
\end{align}
Since we are interested in
the thermal critical coupling $\gTC$,
we set the reference mass to be the critical temperature itself: 
$M(\gLR) = 1/N_t ~ a(\gL^c)$. Inserting this into Eq. (\ref{eq:RV}),
we see that $\gTC$ is implicitly given as:
\begin{align}
R(g_L^c,\gTC) = 1/N_t\ ,
\end{align}
where
we use the following $g_L^c$ from Table \ref{Tab:bc}:
\begin{align}
\gLC
&=
\sqrt{10/\betaLC} 
=
\begin{cases}
1.100\pm 0.004\quad (N_f = 0,\ N_t = 8)\\
1.275\pm 0.040\quad (N_f = 4,\ N_t = 8)\\
1.342\pm 0.032\quad (N_f = 6,\ N_t = 12)\\
1.518\pm 0.021\quad (N_f = 8,\ N_t = 12)
\end{cases}
\ ,\label{eq:gc_best}
\end{align}
Alternative choices corresponding to smaller $N_t$ produce
results for $\gTC$ suffering from the modest scaling 
violations discussed above. 

\begin{figure}
\begin{center}
\includegraphics[width=10.5cm]{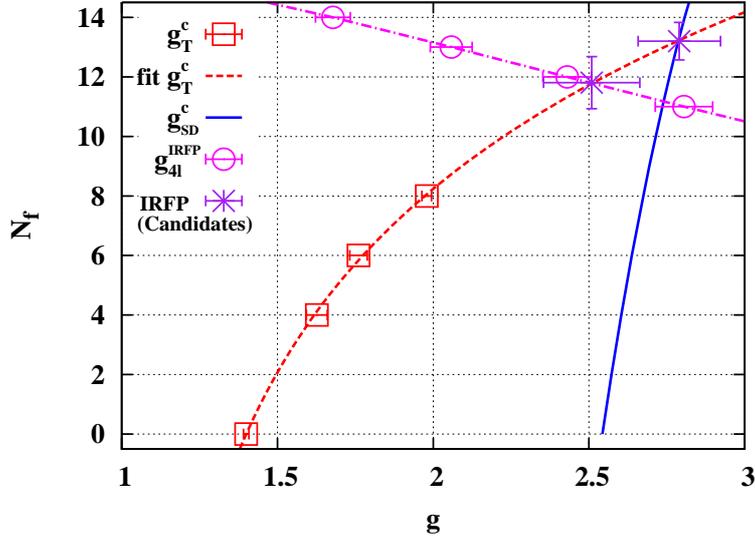}
\caption{
The thermal critical coupling (red $\Box$)
and the fit for them
(dashed red line, with the ansatz Eq.~(\protect\ref{eq:gTC_fit}))
and the values of the zero temperature couplings in the conformal
phase from different estimates, see text for details.
At the critical
number of flavour the thermal critical coupling
should equal the critical
coupling associated with the IRFP.
The procedure is motivated by a recent
study by Shuryak in Ref.~\protect\cite{Liao:2012tw}.}
\label{Fig:gTC}
\end{center}
\end{figure}

The red ($\Box$) symbol in
Fig.~\ref{Fig:gTC} shows $\gTC$ as a function of $N_f$.
We superimpose a fit obtained by using the ansatz proposed in
Ref. \cite{Liao:2012tw}
\begin{align}
{N_f(\gTC) = A\cdot \log~
\bigl[B\cdot(\gTC- \gTC|_{N_f=0}) + 1\bigr]\ .\label{eq:gTC_fit}}
\end{align}
with $A$ and $B$ fit parameters, which describes well the data. 

Since the critical temperature is
zero in the conformal phase,
the thermal critical coupling $\gTC$
should equal a {\em zero temperature} critical coupling $g^c$
when $N_f = N_f^*$.
Of course, $g^c$ is not known exactly
and we have to rely on approximations. 

The first estimate is based on
the best available value $\gSD$
obtained by using
the two-loop Schwinger-Dyson equation \cite{Appelquist:1998rb}.
In this case,
the lower edge of the conformal window $N_f^*$
is defined by the condition $\gTC(N_f^*) = \gSD(N_f^*)$.
In Fig.~\ref{Fig:gTC}  $\gSD$ is plotted as a blue solid line. 
We then estimate the intersection of $\gTC$ and $\gSD$ --
hence the onset of the conformal window
as well as the IRFP coupling at $N_f^*$ -- 
at $(g^*,N_f^*) = (2.79,13.2)\pm (0.13,0.6)$.

One second possibility for estimating $N_f^*$ is the following:
the conformal phase would emerge
when the coupling at IRFP
($g^{\mathrm{IRFP}}$) is not strong enough 
to break  chiral symmetry, {\em i.e.}
$g^{\mathrm{IRFP}} \leq \gTC$.
Here, we utilise the four-loop result for $\gIRFP$
\cite{Ryttov:2012nt} as the best available.
In Fig.~\ref{Fig:gTC}, we show 
$\gIRFP$ as magenta $\bigcirc$,
with superimposed a linear interpolation.
In the plot,
we use the results for $\gIRFP$ in the $\bar{\text{MS}}$ scheme.
The errors are estimated by considering the scheme 
dependence \cite{Ryttov:2012nt}, which turns out to be rather mild
at four loops.
We can then locate the intersection of $\gTC$ and $\gIRFP$
and obtain $(g^*,N_f^*) = (2.51,11.8)\pm (0.15,0.9)$.

Ideally,
the three lines in Fig. \ref{Fig:gTC}
should meet at a (single) IRFP fixed point, if all the
quoted results -- including the analytic ones -- were exact.
Indeed 
the intersections we have estimated are consistent 
within the largish errors.
We then quote the average of the above two estimates
as our final result from this analysis, $N_f^*\sim 12.5\pm 1.6$.

In addition, and on a slightly different aspect, 
we note that $\gTC$ is an increasing function
of $N_f$. This indicates that the quark-gluon plasma
is more strongly coupled at larger $N_f$,
as discussed in Ref.~\cite{Liao:2012tw}.

\subsection{The critical number of flavor
and the  vanishing critical temperature}
\label{subsec:TcM}

Finally, we present an estimate of the onset of the conformal
window which closely follows our previous approach \cite{Miura:2011mc},
based on the analysis of the $N_f$ dependence of
the pseudo-critical temperature in units of
a UV dominated scale.
In this subsection,
we introduce and exploit a new UV reference scale $M$.

\begin{figure}
\begin{center}
\includegraphics[width=10.0cm]{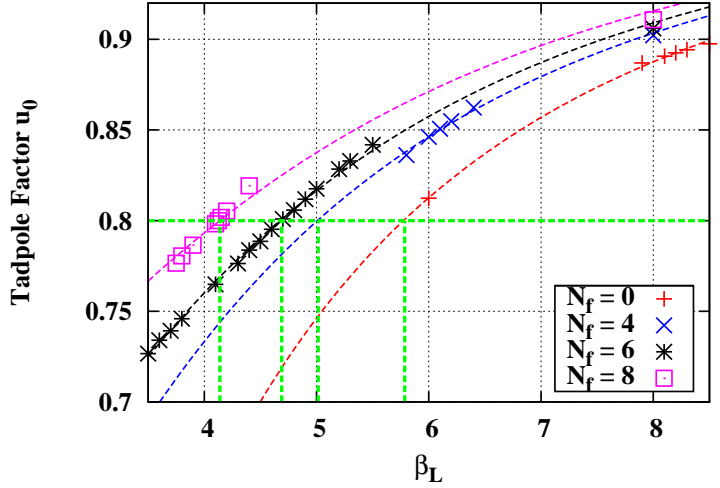}
\caption{
The $\betaL$ dependences of the tadpole factor $u_0$
at zero temperature ($12^4$ lattice volume).
{At each $N_f$,
the dashed line represents the fit for data
with the ansatz $u_0 = 1 - A/(1 + B\cdot\betaL^2)$.}
We consider a constant $u_0$ ({\em e.g.} $u_0=0.8$ in figure),
and read off the corresponding lattice bare couplings $\betaL$,
which are used to define the scale $M$ at each
theory with $N_f$ flavours.}
\label{Fig:u0_const}
\end{center}
\end{figure}

Before going to details, we first explain the basic idea
which follows the FRG analysis by Braun and Gies~\cite{BraunGies}. 
They used the $\tau$ lepton mass $m_{\tau} = 1.777$ (GeV)
as an $N_f$ independent UV reference scale for
theories with any number of flavours.
The initial condition of the renormalisation flow
has been specified via the strong coupling constant
in an $N_f$ independent way:
\begin{align}
\alpha_s(\mu = m_{\tau}) = 0.322\quad
\text{for}\quad {}^{\forall}N_f\ .\label{eq:ini_FRG}
\end{align}
Starting from the common initial condition Eq.~(\ref{eq:ini_FRG}),
 the $N_f$ dependence of the critical temperature $T_c(N_f)$
emerges from 
the 
$N_f$ dependent 
renormalisation flow
at the chiral phase transition scale
$\mu\sim\Lambda_{\mathrm{QCD}}\ll m_{\tau}$.
The $N_f$ dependence of $T_c$
as well as its novel non-analytic behaviour
in the pre-conformal region
becomes free from the choice of
the reference scale~\cite{BraunGies}
by using an $N_f$ independent UV reference scale
much larger than $T_c$.

Ideally,
we would like to 
set our UV scale by measuring on the lattice
some physical quantity insensitive to IR dynamics --
for instance by fixing the value of $\alpha_s$ in the
V-scheme to some appropriate value, as done in
the computation of $r_0$ or variations thereof,
following Ref.~\cite{Sommer:1993ce} and related applications. 
These large scale simulations are now starting \cite{future}.
Here we design a simplified procedure. 

In order to determine the reference coupling $\gLR$
which appears in Eq.~(\ref{eq:RV}),
we utilise our plaquette results  $\langle P \rangle$ (equivalently,
the tadpole factor $u_0 = \langle P \rangle^{1/4}$)
shown in Fig.~\ref{Fig:u0_const}. 

Let us consider a constant $u_0$,
for instance $u_0 = 0.8$ in figure,
and read off the corresponding bare lattice couplings at each $N_f$.
The obtained $\gL(N_f)$ 
is used as a reference coupling $\gLR$
and the corresponding mass scale $M(\gLR)$
is again computed according to Eq.~(\ref{eq:RV}).

Some remarks on the aforementioned scale setting are in order:
First, we recall the scale setting procedure in the potential scheme,
where the measured normalised force $r^2F(r)$
is proportional to the renormalised coupling $\bar{g}$,
and the specification $\bar{g}^2\propto r_X^2F(r_X) = {}^\exists X$
sets a scale $r_X^{-1}$.
In short, we use our $u_0$ (or equivalently plaquettes) 
to define $\bar{g}$,  
and $u_0 = X$ is regarded as the analog of
the potential scheme scale setting.
Second,
in the leading order of the perturbative expansion,
the renormalised coupling is $N_f$ independent, and
proportional to the Wilson loop~\cite{Wong:2005jx} --
a property that we have already exploited in 
the E-scheme calculation.
Hence the use of an $N_f$ independent $u_0$
approximately gives an $N_f$ independent scale setting,
similarly to
the FRG scale setting method Eq.~(\ref{eq:ini_FRG}).
And third, such an $N_f$ independent scale setting
can be performed in a sufficiently UV regime
$T_c(N_f) \ll M(\gLR)$
by adjusting the value of $u_0$ to satisfy
the condition $\gLR \ll \gTC({}^{\forall}N_f)$.

\begin{figure}
\begin{center}
\includegraphics[width=7.5cm,height=5.6cm]{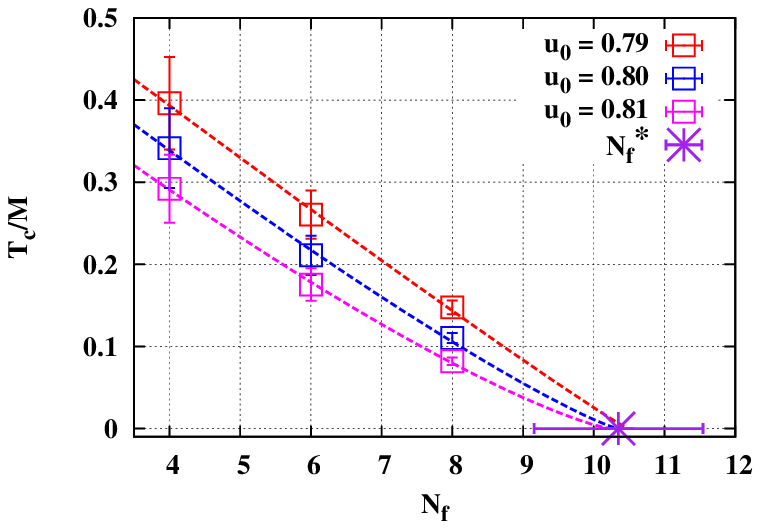}
\includegraphics[width=7.5cm,height=5.6cm]{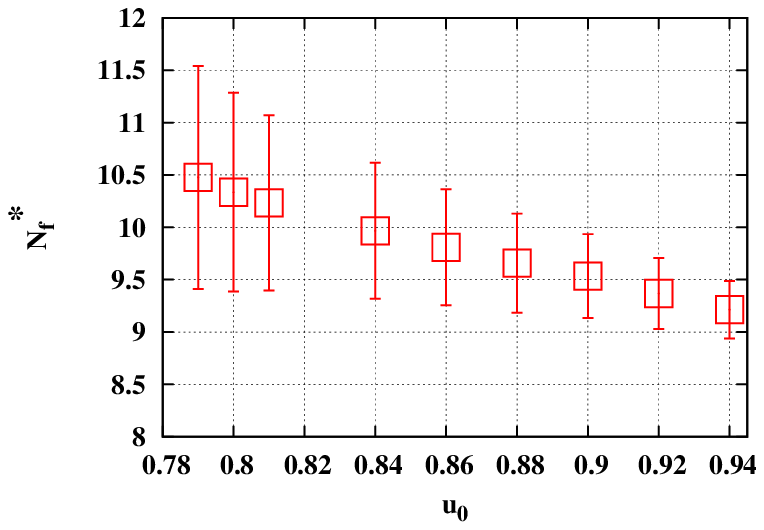}
\caption{
Left:~
The $N_f$ dependence of $T_c/M$ where
$M$ is determined to be a UV scale corresponding to
$u_0=0.79$ (red box),
$0.80$ (blue $\bigcirc$), and
$0.81$ (magenta triangle)
at each theory with $N_f$.
The dashed lines represent fits for data
by assuming the expression Eq.~(\protect\ref{eq:BG_scaling}).
Right: The $u_0$ dependence of $N_f^*$.
The three data in the left side
are determined within the condition $M(\gLR) \lesssim a^{-1}(\gLC)$,
while for the others $M(\gLR)$ exceeds the lattice cutoff.
{This more robust procedure confirms our early results, and should
be ultimately confirmed by use a rigorous lattice scale setting
which is in progress \protect\cite{future}}.}
\label{Fig:TcM}
\end{center}
\end{figure}
Note that the 
coupling at the lattice cutoff $a^{-1}(\gLC)$
is $N_t\gg 1$ times larger than $T_c$.
Then, the scale hierarchy $T_c(N_f) \ll a^{-1}(\gLC(N_f))$
allows us to consider a reference scale
much larger than critical temperature
but smaller than the lattice cutoff
$T_c(N_f) \ll M(\gLR) < a^{-1}(\gLC(N_f))$.
We find that $u_0\sim 0.8$ meets this requirement.

In summary, 
the use of $\gLR$ given by $u_0\sim 0.8$
is analogous to
the FRG scale setting method Eq.~(\ref{eq:ini_FRG}),
and is suitable for studying the vanishing
of the critical temperature
by utilising $T_c/M(\gLR)$.

The left panel of Fig.~\ref{Fig:TcM}
displays the $N_f$ dependence of $T_c/M(\gLR)$
for $u_0 = 0.79$, $0.80$, and $0.81$.
Fitting the data points for $T_c/M(\gLR)$ at $N_f \geq 4$
by using the FRG motivated ansatz,
\begin{align}
T_c= K|N_f^* - N_f|^{(-2b_0^2/b_1)(N_f^*)}
\ ,\label{eq:BG_scaling}
\end{align}
where $b_{0,1}$ has been defined in Eqs.~(\ref{eq:b0}) and (\ref{eq:b1}),
the lower edge of the conformal window is estimated as:
$N_f^* = 10.48 \pm 1.01$ ($u_0 = 0.79$),
$N_f^* = 10.34 \pm 0.88$ ($u_0 = 0.80$),
$N_f^* = 10.23 \pm 0.80$ ($u_0 = 0.81$).
The error-bars involve both fit errors and
statistical errors of data.

We have further investigated the stability against
different choices of $u_0$:
As shown in the right panel of Fig. \ref{Fig:TcM},
$N_f^*$ is relatively stable
within the range $0.79\leq u_0\leq 0.94$.
The scale cannot be pushed further towards
the UV because of discritization errors. 
On the other hand, 
a small $u_0 \lesssim 0.7$ leads to $M(\gLR)\sim T_c$ or smaller.
In such a case,
the reference scale $M(\gLR)$ 
is affected by infra-red physics and cannot
be used to study the vanishing of $T_c$.
Despite these limitations,
the window of relative stability is however reasonably large,
and suffices to define
an average value for $N_f^*$.
We quote the average among
the three results obtained for $u_0=(0.79,0.80,0.81)$,
i.e. $N_f^* = 10.4 \pm 1.2$.


\section{Summary}\label{sec:sum}
We have investigated the phase transition (crossover)
at finite temperature $T$
in colour SU$(3)$ QCD-like theories
with various number of flavours
$N_f = 0$ (quenched), $4$, $6$, and $8$
by using lattice Monte Carlo simulations.
We have used a single bare fermion mass $ma = 0.02$.
For all number of flavours,
we have used the Asqtad action with a one-loop Symanzik
and tadpole improved gauge action.
The main focus in this paper is to investigate
the chiral crossover at finite $T$ as a function of $N_f$,
and discuss the possible implication for the
(pre-)conformal dynamics at large $N_f$.
In Eq.~(\ref{eq:Nf_IRFP}),
we provide the summary for the number of flavour $N_f^*$
where the conformal window would emerge.

The observables in our simulations were
the chiral condensate, the Polyakov loop,
and their susceptibilities
for various lattice couplings $\betaL$,
lattice sizes, and the number of flavours $N_f$.
We have collected the (pseudo) critical lattice coupling
$\betaLC$ as a function of ($N_f,N_t$).
Table \ref{Tab:bc} provides the summary for $\betaLC$.
Our $\betaLC$ are consistent
with enhanced fermionic screening at larger $N_f$.
The use of several $N_t$ allows us to
study the asymptotic scaling of the critical temperature.
Further, by  utilising $\betaLC$,
we have discussed a possible implication for
the (pre-)conformal dynamics at large $N_f$.

We have estimated the $N_f^*$ from the vanishing thermal scaling
by extrapolating our critical couplings $\gLC$ to
larger $N_f$ . This gives $N_f^*\sim 11.1\pm 1.6$.
We have extracted
a typical interaction strength $\gTC$
at the scale of critical temperature $T_c$
by utilising our $\gLC$ and the two-loop beta-function,
and compared $\gTC$ to
the zero temperature critical couplings ($\gSD$)
estimated by the two-loop Schwinger-Dyson equation
\cite{Appelquist:1998rb}
as well as the IRFP position
($\gIRFP$)
of the four-loop beta-function \cite{Ryttov:2012nt}.
The coincidence between $\gTC$ and $\gSD$ or
$\gIRFP$ indicates
the vanishing critical temperature
with the emergence of the conformal phase.
Based on this reasoning,
we have estimated the onset of the conformal window
as $N_f^*\sim 12.5\pm 1.6$.
We have also confirmed
the increasing of $\gTC$ at larger $N_f$
which has been discussed in Ref.~\cite{Liao:2012tw}
and indicates
more strongly interacting non-Abelian plasma at larger $N_f$.

Further, we have examined the $N_f$ dependence of $T_c/M$
by introducing a UV $N_f$ independent reference scale $M$
which is determined by utilising the tadpole factor $u_0$
in analogous ways to the potential scheme scale setting.
Then, $T_c/M$ turns out to be a decreasing function of $N_f$
consistently to the FRG observations~\cite{BraunGies},
and the vanishing $T_c/M$ indicates the emergence of the conformal window
around $N_f^* \sim 10.4 \pm 1.2$.

As a future perspective, we plan to
perform more rigorous scale settings, by
exploiting state-of-art measurements of lattice potential.
It is also mandatory to investigate the chiral limit
and the thermodynamic limit at large $N_f$.
This, together with a more extended set of flavour
numbers, will allow a quantitative analysis of the critical
behaviour in the vicinity of the conformal IR fixed point.
We expect that
our {\em thermodynamic} lattice study for
the large $N_f$ non-Abelian gauge theory
plays an important role as a new connection
between the lattice and the Gauge/Gravity duality
\cite{Gursoy:2010fj,Alho:2012mh}.

\section*{Acknowledgements}
We enjoyed discussing these topics with George Fleming,
Edward Shuryak, Anna Hasenfratz, Philippe de Forcrand, and Marc Wagner. 
We also wish to thank Carleton DeTar 
and Urs Heller for sharing their notes on
the normalization of chiral observables in the MILC code.
This project is one
step in our study of the phase diagram of strong interactions, 
and we warmly thank Elisabetta Pallante, 
Albert Deuzeman and Tiago Nunes da Silva for many interesting
conversations and a pleasant collaboration. 
We also thank them
for granting access to some of $u_0$ data at $N_f = 8$ 
used in this study,
and in particular Tiago Nunes da Silva for related communications.
This work was in part based on the MILC Collaboration's public
lattice gauge theory code~\cite{MILC}.
The numerical calculations were carried out on the
IBM-SP6 and BG/P at CINECA,
Italian-Grid-Infrastructures in Italy,
and the Hitachi SR-16000 at YITP, Kyoto University in Japan.
\appendix
\section{Summary table for simulation results}\label{app:sum_table}

In this appendix,
we summarise our results and the parameters
used in the simulations and analyses.
We have used an improved version of the staggered action,
the Asqtad action~\cite{Lepage:1998vj},
with a one-loop Symanzik \cite{Bernard:2006nj,LuscherWeisz}
and tadpole~\cite{LM1985} improved gauge action.
All our simulations used 
the same  bare fermion mass $ma = 0.02$.

In Tables~\ref{tab:Nf0_Nt4} - \ref{tab:Nf8_Nt8}
we will quote the
lattice bare couplings $\betaL = 10/\gL^2$,
tadpole factors $u_0$,
step lengths for a single trajectory
$\Delta\tau=\delta \tau\times 20$,
the number of total trajectories $n_{\mathrm{traj}}$,
the number of thermalized trajectories
$n_{\mathrm{ave}}$ which have been used to evaluate
ensemble averages,
bin-sizes for jackknife error estimates $s_{\mathrm{bin}}$,
ensemble averages for the chiral condensate
($2\PBP$ with the definition Eq.~(\ref{eq:PBP}))
the Polyakov loop
($\mathrm{Re}~L$ or $|L|$ with the definition Eq.~(\ref{eq:PLOOP})),
and the chiral susceptibility
($\chi_{\sigma}$, defined in Eq.~(\ref{eq:sus_sig})) and/or
the susceptibility ($\chi_{|L|}$) associated with
the absolute value of Polyakov loop.

Here are several technical comments for the symbols:
In the second column in each table,
some values for the tadpole factors $u_0$
have been estimated by using the fit with the ansatz
$u_0 = 1.0 - A/(1.0 + B~\betaL^2)$,
and are shown inside of the parentheses.
Some of the total trajectories in the third column
have a symbol $+$, for which
the configurations obtained in the other simulations
has been utilised as the inputs.
The Monte Carlo step $\Delta\tau$
in the fourth column 
{has been adjusted} 
to give about $75 - 80$ percent Metropolis acceptances
in the pre-thermals domain.
In the fifth column,
some entries have a symbol $\ast$,
which indicates that additional simulations
would be preferable to confirm the thermalization
though the trajectories seem
to have reached a  stable domain.
In the seventh column,
a pre-factor $2$ appears in front of $\PBP$
following the difference between the  
the MILC code convention and
the standard definition Eq.~(\ref{eq:PBP}).

\begin{table*}[th]
\caption{Summary table for $N_f = 0$
with the use of the lattice volume $16^3\times 4$.
}\label{tab:Nf0_Nt4}
\begin{center}
\begin{tabular}{l|llllllll}
\hline\hline
$\betaL$ & $u_0$ & $n_{\mathrm{traj}}$ & $\Delta\tau$ & $n_{\mathrm{ave}}$ & $s_{\mathrm{bin}}$ & $2\PBP$ & $|L|$ & $\chi_{|L|}$ \\
\hline
$6.80$ & $(0.864)$ & $4000$ & $0.60$ & $2000$ & $200$ & $0.575251(110)$ & $0.024(1)$ & $0.6\pm 0.04$\\
$7.00$ & $(0.868)$ & $4000$ & $0.60$ & $2000$ & $200$ & $0.559423(113)$ & $0.032(2)$ & $1.10\pm 0.11$\\
$7.10$ & $(0.871)$ & $4000$ & $0.60$ & $2000$ & $400$ & $0.552847(121)$ & $0.038(7)$ & $2.29\pm 1.10$\\
$7.15$ & $(0.8715)$ & $3000$ & $0.60$ & $2000$ & $250$ & $0.549450(102)$ & $0.050(3)$ & $2.27\pm 0.28$\\
$7.20$ & $(0.872)$ & $4000$ & $0.60$ & $2000$ & $400$ & $0.546144(63)$ & $0.052(7)$ & $2.99\pm 0.41$\\
$7.25$ & $(0.8735)$ & $3000$ & $0.60$ & $1500$ & $250$ & $0.543789(46)$ & $0.051(4)$ & $2.06\pm 0.32$\\
$7.30$ & $(0.875)$ & $4000$ & $0.60$ & $1500$ & $250$ & $0.541361(119)$ & $0.057(10)$ & $4.75\pm 0.65$\\
$7.35$ & $(0.876)$ & $12000$ & $0.70$ & $6000$ & $1200$ & $0.538570(127)$ & $0.273(29)$ & $29.34\pm 10.51$\\
$7.40$ & $(0.877)$ & $12000$ & $0.68$ & $6000$ & $1200$ & $0.535896(83)$ & $0.423(7)$ & $5.43\pm 0.90$\\
$7.45$ & $0.878$ & $8000$ & $0.68$ & $4000$ & $500$ & $0.531813(120)$ & $0.395(10)$ & $9.59\pm 1.83$\\
$7.50$ & $(0.879)$ & $4000$ & $0.60$ & $2000$ & $400$ & $0.529767(146)$ & $0.440(14)$ & $7.30\pm 0.66$\\
$8.00$ & $(0.889)$ &$1000$ & $0.60$ & $500$ & $100$ & $0.516562(107)$ & $0.702(7)$ & $2.51\pm 0.69$\\
\hline\hline
\end{tabular}
\end{center}
\end{table*}

\begin{table*}[th]
\caption{Summary table for $N_f = 0$
with the use of the lattice volume $16^3\times 6$.
In order to find the suitable $\Delta \tau$,
we performed some preceded simulations
with the order of $5000$ trajectories,
and their output configurations
have been used as inputs to get the results in this table.
}\label{tab:Nf0_Nt6}
\begin{center}
\begin{tabular}{l|llllllll}
\hline\hline
$\betaL$ & $u_0$ &$n_{\mathrm{traj}}$ & $\Delta\tau$ & $n_{\mathrm{ave}}$ & $s_{\mathrm{bin}}$ & $2\PBP$ & $|L|$ & $\chi_{|L|}$ \\
\hline
$7.50$ & $(0.879)$ & $4000+$ & $0.68$ & $600$ & $100$ & $0.532814(60)$ & $0.0278(018)$ & $0.77\pm 0.08$\\
$7.60$ & $(0.881)$ & $8000+$ & $0.68$ & $1000$ & $200$ & $0.529409(42)$ & $0.0372(018)$ & $1.72\pm 0.32$\\
$7.70$ & $(0.883)$ & $12000+$ & $0.68$ & $2000$ & $400$ & $0.526335(44)$ & $0.0652(086)$ & $5.34\pm 0.75$\\
$7.80$ & $(0.885)$ & $18000+$ & $0.68$ & $2000$ & $500$ & $0.523805(11)$ & $0.0656(107)$ & $4.44\pm 1.41$\\
$7.90$ & $0.886994(54)$ & $10000+$ & $0.68$ & $2000$ & $400$ & $0.521344(35)$ & $0.1950(157)$ & $8.20\pm 3.19$\\
$8.00$ & $0.889$ & $6000+$ & $0.68$ & $900$ & $100$ & $0.519265(44)$ & $0.2538(079)$ & $5.08\pm 0.51$\\
$8.10$ & $0.890728(45)$ & $6000+$ & $0.68$ & $1000$ & $250$ & $0.517268(47)$ & $0.2838(085)$ & $4.48\pm 0.51$\\
$8.20$ & $0.892579(35)$ & $6000+$ & $0.68$ & $1000$ & $125$ & $0.515598(21)$ & $0.3255(042)$ & $3.43\pm 0.28$\\
$8.30$ & $0.894318(46)$ & $4000+$ & $0.64$ & $600$ & $100$ & $0.514019(24)$ & $0.3514(061)$ & $3.93\pm 0.42$\\
$8.40$ & $(0.896)$ & $4000+$ & $0.64$ & $600$ & $120$ & $0.512534(31)$ & $0.3806(049)$ & $3.43\pm 0.55$\\
\hline\hline
\end{tabular}
\end{center}
\end{table*}

\begin{table*}[th]
\caption{Summary table for $N_f = 0$
with the use of the lattice volume $24^3\times 8$.
}\label{tab:Nf0_Nt8}
\begin{center}
\begin{tabular}{l|llllllll}
\hline\hline
$\betaL$ & $u_0$ & $n_{\mathrm{traj}}$ & $\Delta\tau$ & $n_{\mathrm{ave}}$ & $s_{\mathrm{bin}}$ & $2\PBP$ & $|L|$ & $\chi_{|L|}$ \\
\hline
$7.80$ & $(0.885)$ & $4800$ & $0.52$ & $1000$ & $200$ & $0.523836(26)$ & $0.0107(5)$ & $0.37\pm 0.02$\\
$8.00$ & $(0.889)$ & $3100$ & $0.56$ & $2000$ & $400$ & $0.519532(22)$ & $0.0139(10)$ & $0.68\pm 0.03$\\
$8.10$ & $0.890728(45)$ & $12000$ & $0.54$ & $1000$ & $250$ & $0.517575(22)$ & $0.0175(21)$ & $1.00\pm 0.09$\\
$8.15$ & $(0.892)$ & $9600$ & $0.56$ & $4000$ & $800$ & $0.516582(15)$ & $0.0566(64)$ & $4.31\pm 1.06$\\
$8.20$ & $0.892579(35)$ & $8800$ & $0.52$ & $2000$ & $400$ & $0.515820(23)$ & $0.0317(32)$ & $2.19\pm 0.45$\\
$8.25$ & $0.893$ & $7200$ & $0.54$ & $3000$ & $600$ & $0.514974(12)$ & $0.0770(71)$ & $4.89\pm 1.24$\\
$8.30$ & $0.894318(46)$ & $8700$ & $0.56$ & $3000$ & $600$ & $0.514153(9)$ & $0.1158(43)$ & $3.32\pm 0.94$\\
$8.40$ & $(0.896)$ & $4800$ & $0.56$ & $1500$ & $300$ & $0.512647(12)$ & $0.1623(36)$ & $2.23\pm 0.47$\\
$8.60$ & $(0.899)$ & $4000$ & $0.56$ & $1500$ & $300$ & $0.509860(4)$ & $0.1807(45)$ & $2.63\pm 0.79$\\
$8.80$ & $(0.900)$ & $4000$ & $0.56$ & $1500$ & $300$ & $0.505961(5)$ & $0.2160(39)$ & $3.53\pm 0.42$\\
$9.00$ & $(0.900)$ & $4000$ & $0.56$ & $1500$ & $300$ & $0.501632(5)$ & $0.2587(33)$ & $2.41\pm 0.56$\\
\hline\hline
\end{tabular}
\end{center}
\end{table*}

\begin{table*}[th]
\caption{Summary table for $N_f = 4$
with the use of the lattice volume $16^3\times 4$.
}\label{tab:Nf4_Nt4}
\begin{center}
\begin{tabular}{l|llllllll}
\hline\hline
$\betaL$ & $u_0$ & $n_{\mathrm{traj}}$ & $\Delta\tau$ & $n_{\mathrm{ave}}$ & $s_{\mathrm{bin}}$ & $2\PBP$ & $a^2\chi_{\sigma}$ & $\mathrm{Re}~L$\\
\hline
$4.80$ & $(0.774)$ & $3000$ & $0.12$ & $1000$ & $250$ & $0.5614(15)$ & $0.356(17)$ & $0.0053(21)$\\
$5.10$ & $(0.796)$ & $4000$ & $0.12$ & $1500$ & $250$ & $0.4733(13)$ & $0.552(17)$ & $0.0141(16)$\\
$5.30$ & $(0.809)$ & $4000+$ & $0.14$ & $1500$ & $250$ & $0.4001(13)$ & $0.688(22)$ & $0.0256(30)$\\
$5.40$ & $(0.815)$ & $6000$ & $0.14$ & $1500$ & $300$ & $0.3485(26)$ & $0.934(49)$ & $0.0406(19)$\\
$5.45$ & $(0.818)$ & $5000+$ & $0.14$ & $3000$ & $500$ & $0.3132(26)$ & $1.090(77)$ & $0.0579(14)$\\
$5.50$ & $(0.821)$ & $4000+$ & $0.14$ & $1500$ & $250$ & $0.2893(11)$ & $0.993(24)$ & $0.0657(15)$\\
$5.55$ & $(0.824)$ & $5000+$ & $0.16$ & $2000$ & $250$ & $0.2443(16)$ & $1.281(48)$ & $0.0917(23)$\\
$5.60$ & $(0.826)$ & $7000$ & $0.16$ & $2000$ & $500$ & $0.1814(25)$ & $1.746(103)$ & $0.1356(35)$\\
$5.65$ & $(0.829)$ & $5000+$ & $0.18$ & $2000$ & $400$ & $0.1117(26)$ & $1.934(97)$ & $0.1953(41)$\\
$5.70$ & $(0.831)$ & $4000$ & $0.20$ & $1500$ & $250$ & $0.0859(14)$ & $1.747(15)$ & $0.2231(27)$\\
$5.80$ & $0.836159(98)$ & $4000+$ & $0.24$ & $1500$ & $250$ & $0.0610(7)$ & $1.425(8)$ & $0.2590(23)$\\
$6.00$ & $0.846224(81)$ & $3000$ & $0.28$ & $1500$ & $150$ & $0.0391(1)$ & $0.964(2)$ & $0.3519(28)$\\
\hline\hline
\end{tabular}
\end{center}
\end{table*}

\begin{table*}[th]
\caption{Summary table for $N_f = 4$
with the use of the lattice volume $16^3\times 6$.}
\label{tab:Nf4_Nt6}
\begin{center}
\begin{tabular}{l|llllllll}
\hline\hline
$\betaL$ & $u_0$ & $n_{\mathrm{traj}}$ & $\Delta\tau$ & $n_{\mathrm{ave}}$ & $s_{\mathrm{bin}}$ & $2\PBP$ & $a^2\chi_{\sigma}$ & $\mathrm{Re}~L$\\
\hline
$5.70$ & $(0.831)$ & $3500$ & $0.12$ & $1000$ & $200$ & $0.2311(8)$ & $0.948(8)$ & $0.0025(15)$\\
$5.80$ & $0.836159(98)$ & $4500$ & $0.16$ & $1000$ & $200$ & $0.1772(13)$ & $1.102(29)$ & $0.0125(10)$\\
$5.85$ & $(0.8385)$ & $6000$ & $0.16$ & $1000$ & $200$ & $0.1617(9)$ & $1.057(10)$ & $0.0094(19)$\\
$5.90$ & $(0.841)$ & $5500$ & $0.16$ & $3000$ & $750$ & $0.1338(28)$ & $1.270(94)$ & $0.0180(32)$\\
$5.95$ & $(0.8436)$ & $6000$ & $0.16$ & $2000$ & $500$ & $0.1000(14)$ & $1.311(30)$ & $0.0394(18)$\\
$6.00$ & $0.846224(81)$ & $8000$ & $0.18$ & $4000$ & $800$ & $0.0831(25)$ & $1.499(98)$ & $0.0463(33)$\\
$6.10$ & $0.850711(61)$ & $2500$ & $0.20$ & $1000$ & $200$ & $0.0503(4)$ & $1.152(12)$ & $0.0854(26)$\\
$6.20$ & $0.854878(58)$ & $1600$ & $0.20$ & $600$ & $100$ & $0.0413(3)$ & $1.003(7)$ & $0.1081(20)$\\
$6.30$ & $(0.859)$ & $1500$ & $0.24$ & $500$ & $85$ & $0.0366(2)$ & $0.902(5)$ & $0.1198(21)$\\
$6.40$ & $0.862358(65)$ & $1200$ & $0.24$ & $400$ & $80$ & $0.0330(2)$ & $0.816(33)$ & $0.1464(28)$\\
\hline\hline
\end{tabular}
\end{center}
\end{table*}

\begin{table*}[th]
\caption{Summary table for $N_f = 4$
with the use of lattice volume $24^3\times 8$.
For the simulations with $\betaL = 6.5$ and $6.8$,
the Monte Carlo step $\Delta\tau$ must be set to be
smaller than the presented values in the early stage
of the molecular dynamics evolutions
to avoid low Metropolis acceptances.
}\label{tab:Nf4_Nt8}
\begin{center}
\begin{tabular}{l|llllllll}
\hline\hline
$\betaL$ & $u_0$ & $n_{\mathrm{traj}}$ & $\Delta\tau$ & $n_{\mathrm{ave}}$ & $s_{\mathrm{bin}}$ & $2\PBP$ & $a^2\chi_{\sigma}$ & $\mathrm{Re}~L$\\
\hline
$5.60$ & $(0.826)$ & $1500+$ & $0.12$ & $400 $ & $100$ & $0.2829(8) $ & $0.861(23)$ & $-0.0004(23)$\\
$5.80$ & $0.836159(98)$ & $2100$ & $0.14$ & $600 $ & $75 $ & $0.1823(3) $ & $0.965(19)$ & $0.0009(8)$\\
$5.90$ & $0.841$ & $1500+$ & $0.14$ & $1000 $ & $200 $ & $0.1453(5) $ & $1.017(12)$ & $0.0014(5)$\\
$5.95$ & $(0.8436)$ & $4000+$ & $0.14$ & $1600 $ & $400 $ & $0.1268(10)$ & $1.176(26)$ & $0.0015(6)$\\
$6.00$ & $0.846224(81)$ & $7000$ & $0.16$ & $2500 $ & $625 $ & $0.1098(10)$ & $1.113(24)$ & $0.0031(3)$\\
$6.05$ & $(0.848)$ & $4800+$ & $0.16$ & $3500\ast $ & $875 $ & $0.0973(11)$ & $1.107(9) $ & $0.0046(5)$\\
$6.10$ & $0.850711(61)$ & $5400$ & $0.16$ & $2000 $ & $500 $ & $0.0828(7) $ & $1.066(23)$ & $0.0068(4)$\\
$6.15$ & $(0.853)$ & $2000+$ & $0.16$ & $800 $ & $100 $ & $0.0690(3) $ & $1.046(10)$ & $0.0122(9)$\\
$6.20$ & $0.854878(58)$ & $2500$ & $0.18$ & $800 $ & $114 $ & $0.0584(3) $ & $1.046(9) $ & $0.0165(4)$\\
$6.25$ & $(0.8567)$ & $4000+$ & $0.18$ & $2000 $ & $400 $ & $0.0518(3) $ & $1.035(13)$ & $0.0209(13)$\\
$6.30$ & $(0.859)$ & $2540$ & $0.22$ & $1000 $ & $200 $ & $0.0435(3) $ & $1.006(7) $ & $0.0280(13)$\\
$6.35$ & $(0.8608)$ & $1590+$ & $0.22$ & $1200 $ & $240 $ & $0.0401(4) $ & $0.962(5) $ & $0.0346(10)$\\
$6.40$ & $0.862358(65)$ & $3500$ & $0.24$ & $1500 $ & $375 $ & $0.0384(3) $ & $0.917(6) $ & $0.0378(8)$\\
$6.45$ & $(0.8647)$ & $1590+$ & $0.26$ & $800 $ & $160 $ & $0.0358(1) $ & $0.871(2) $ & $0.0425(13)$\\
$6.50$ & $(0.867)$ & $2350$ & $0.24$ & $700 $ & $100 $ & $0.0340(1) $ & $0.830(2) $ & $0.0501(12)$\\
$6.80$ & $(0.880)$ & $3300$ & $0.40$ & $1000 $ & $200 $ & $0.0284(0) $ & $0.707(1) $ & $0.0817(14)$\\
\hline\hline
\end{tabular}
\end{center}
\end{table*}

\begin{table*}[th]
\caption{Summary table for $N_f = 6$
with the use of the lattice volume $16^3\times 4$.}
\label{tab:Nf6_Nt4}
\begin{center}
\begin{tabular}{l|llllllll}
\hline\hline
$\betaL$ & $u_0$ & $n_{\mathrm{traj}}$ & $\Delta\tau$ & $n_{\mathrm{ave}}$ & $s_{\mathrm{bin}}$ & $2\PBP$ & $a^2\chi_{\sigma}$ & $\mathrm{Re}~L$\\
\hline
$4.00$ & $(0.759875)$ & $2000$ & $0.10$ & $1000$ & $300$ & $0.5829(1)$ & $0.304(5)$ & $0.0011(18)$\\
$4.30$ & $0.776360(228)$ & $3000$ & $0.10$ & $1500$ & $300$ & $0.4957(21)$ & $0.585(48)$ & $0.0151(36)$\\
$4.40$ & $0.783740(347)$ & $3000$ & $0.10$ & $1500$ & $300$ & $0.4565(11)$ & $0.615(7)$ & $0.0204(7)$\\
$4.50$ & $0.788558(123)$ & $4000$ & $0.10$ & $1000$ & $300$ & $0.4077(24)$ & $0.778(45)$ & $0.0349(48)$\\
$4.60$ & $0.795206(97)$ & $5000$ & $0.10$ & $2000$ & $300$ & $0.3476(15)$ & $0.982(27)$ & $0.0557(14)$\\
$4.65$ & $(0.798)$ & $4000+$ & $0.16$ & $2000$ & $1000$ & $0.2667(52)$ & $2.609(451)$ & $0.0940(37)$\\
$4.70$ & $0.800839(277)$ & $4000$ & $0.24$ & $1000$ & $300$ & $0.1244(22)$ & $2.206(38)$ & $0.1797(23)$\\
$4.80$ & $0.805839(364)$ & $4000$ & $0.10$ & $1000$ & $300$ & $0.0822(12)$ & $1.831(15)$ & $0.2195(14)$\\
$5.00$ & $0.817551(86)$ & $2000$ & $0.10$ & $1000$ & $300$ & $0.0521(6)$ & $1.271(8)$ & $0.2850(26)$\\
$5.20$ & $0.828421(354)$ & $2000$ & $0.10$ & $1000$ & $300$ & $0.0418(3)$ & $1.035(8)$ & $0.3300(57)$\\
$5.50$ & $0.841873(95)$ & $2000$ & $0.10$ & $1000$ & $300$ & $0.0325(0)$ & $0.810(1)$ & $0.4070(41)$\\
\hline\hline
\end{tabular}
\end{center}
\end{table*}

\begin{table*}[th]
\caption{Summary table for $N_f = 6$
with the use of the lattice volume $16^3\times 6$.}
\label{tab:Nf6_Nt6}
\begin{center}
\begin{tabular}{l|llllllll}
\hline\hline
$\betaL$ & $u_0$ & $n_{\mathrm{traj}}$ & $\Delta\tau$ & $n_{\mathrm{ave}}$ & $s_{\mathrm{bin}}$ & $2\PBP$ & $a^2\chi_{\sigma}$ & $\mathrm{Re}~L$\\
\hline
$3.60$ & $0.734031(104)$ & $1500$ & $0.16$ & $1000$ & $250$ & $0.6622(1)$ & $0.14(1)$ & $-0.0034(6)$\\
$4.00$ & $(0.759875)$ & $1500$ & $0.16$ & $1000$ & $250$ & $0.5923(5)$ & $0.30(1)$ & $-0.0041(4)$\\
$4.20$ & $(0.773118)$ & $3000$ & $0.16$ & $1000$ & $250$ & $0.5422(7)$ & $0.55(14)$ & $-0.0021(23)$\\
$4.40$ & $0.783740(347)$ & $3000$ & $0.16$ & $1000$ & $250$ & $0.4808(0)$ & $0.72(16)$ & $-0.0050(4)$\\
$4.50$ & $0.788558(123)$ & $3000$ & $0.16$ & $1000$ & $250$ & $0.4363(2)$ & $1.12(26)$ & $-0.0032(11)$\\
$4.60$ & $0.795206(97)$ & $3000$ & $0.18$ & $1000$ & $250$ & $0.3947(20)$ & $0.96(17)$ & $-0.0008(4)$\\
$4.70$ & $0.800839(277)$ & $3000$ & $0.20$ & $1000$ & $250$ & $0.3418(10)$ & $1.12(17)$ & $-0.0007(7)$\\
$4.80$ & $0.805839(364)$ & $3000$ & $0.24$ & $1000$ & $250$ & $0.2864(42)$ & $1.43(18)$ & $0.0028(3)$\\
$4.90$ & $0.811809(354)$ & $3000$ & $0.20$ & $1000$ & $250$ & $0.2231(34)$ & $1.64(15)$ & $0.0058(14)$\\
$4.95$ & $(0.81466)$ & $4000$ & $0.24$ & $1000$ & $250$ & $0.1935(17)$ & $1.37(5)$ & $0.0091(1)$\\
$5.00$ & $0.817551(86)$ & $7000$ & $0.20$ & $2000$ & $500$ & $0.1644(28)$ & $1.66(8)$ & $0.0137(10)$\\
$5.05$ & $(0.8196)$ & $6000+$ & $0.24$ & $1000$ & $250$ & $0.1072(63)$ & $1.98(19)$ & $0.0388(32)$\\
$5.10$ & $(0.821629)$ & $3000$ & $0.20$ & $1000$ & $250$ & $0.0769(21)$ & $1.51(10)$ & $0.0566(17)$\\
$5.20$ & $0.828421(354)$ & $3000$ & $0.24$ & $1000$ & $250$ & $0.0581(2)$ & $1.38(1)$ & $0.0710(2)$\\
$5.30$ & $0.832865(89)$ & $3000$ & $0.20$ & $1000$ & $250$ & $0.0489(0)$ & $1.13(6)$ & $0.0850(18)$\\
$5.50$ & $0.841873(95)$ & $3000$ & $0.28$ & $1000$ & $250$ & $0.0393(0)$ & $0.97(0)$ & $0.1138(3)$\\
\hline\hline
\end{tabular}
\end{center}
\end{table*}

\begin{table*}[th]
\caption{Summary table for $N_f = 6$
with the use of the lattice volume $24^3\times 8$.}
\label{tab:Nf6_Nt8}
\begin{center}
\begin{tabular}{l|llllllll}
\hline\hline
$\betaL$ & $u_0$ & $n_{\mathrm{traj}}$ & $\Delta\tau$ & $n_{\mathrm{ave}}$ & $s_{\mathrm{bin}}$ & $2\PBP$ & $a^2\chi_{\sigma}$ & $\mathrm{Re}~L$\\
\hline
$4.60$ & $0.795206(97)$ & $1000$ & $0.20$ & $500$ & $100$ & $0.3985(4)$ & $0.68(4)$ & $-0.0014(11)$\\
$4.80$ & $0.805839(364)$ & $1000$ & $0.20$ & $500$ & $100$ & $0.2929(15)$ & $1.08(5)$ & $-0.0008(0)$\\
$4.90$ & $0.811809(354)$ & $1210$ & $0.20$ & $500$ & $50$ & $0.2335(16)$ & $1.22(12)$ & $0.0001(3)$\\
$5.00$ & $0.817551(86)$ & $3500+$ & $0.24$ & $1000$ & $200$ & $0.1840(7)$ & $1.31(2)$ & $-0.0001(4)$\\
$5.10$ & $(0.821629)$ & $3000$ & $0.24$ & $1500$ & $300$ & $0.1357(8)$ & $1.30(3)$ & $0.0022(4)$\\
$5.15$ & $(0.825)$ & $5000+$ & $0.28$ & $2000$ & $400$ & $0.1164(10)$ & $1.37(6)$ & $0.0037(3)$\\
$5.20$ & $0.828421(354)$ & $5000$ & $0.28$ & $2500\ast $ & $500$ & $0.0959(18)$ & $1.66(14)$ & $0.0067(5)$\\
$5.25$ & $(0.8306)$ & $5000+$ & $0.32$ & $3000$ & $500$ & $0.0732(9)$ & $1.50(7)$ & $0.0126(7)$\\
$5.30$ & $0.832865(89)$ & $4000$ & $0.32$ & $1500$ & $300$ & $0.0606(5)$ & $1.36(1)$ & $0.0177(4)$\\
$5.50$ & $0.841873(95)$ & $1500$ & $0.32$ & $500$ & $125$ & $0.0441(1)$ & $1.07(0)$ & $0.0301(7)$\\
\hline\hline
\end{tabular}
\end{center}
\end{table*}

\begin{table*}[th]
\caption{Summary table for $N_f = 6$
with the use of the lattice volume $24^3\times 12$.}
\label{tab:Nf6_Nt12}
\begin{center}
\begin{tabular}{l|llllllll}
\hline\hline
$\betaL$ & $u_0$ & $n_{\mathrm{traj}}$ & $\Delta\tau$ & $n_{\mathrm{ave}}$ & $s_{\mathrm{bin}}$ & $2\PBP$ & $a^2\chi_{\sigma}$ & $\mathrm{Re}~L$\\
\hline
$4.70$ & $0.800839(277)$ & $1000$ & $0.20$ & $400$ & $100$ & $0.3493(6)$ & $0.753(20)$ & $-0.000188(377)$\\
$4.80$ & $0.805839(364)$ & $1000$ & $0.20$ & $400$ & $100$ & $0.2961(4)$ & $0.927(35)$ & $-0.000096(510)$\\
$4.90$ & $0.811809(354)$ & $1000$ & $0.20$ & $400$ & $100$ & $0.2330(7)$ & $1.200(70)$ & $-0.000344(692)$\\
$5.00$ & $0.817551(86)$ & $1500$ & $0.20$ & $1500$ & $500$ & $0.1838(9)$ & $1.356(50)$ & $-0.000616(710)$\\
$5.10$ & $(0.821629)$ & $1500$ & $0.20$ & $900$ & $300$ & $0.1420(12)$ & $1.269(63)$ & $-0.000297(360)$\\
$5.20$ & $0.828421(354)$ & $1500$ & $0.20$ & $600$ & $200$ & $0.1065(2)$ & $1.172(23)$ & $0.000010(144)$\\
$5.30$ & $0.832865(89)$ & $1500$ & $0.24$ & $900$ & $225$ & $0.0838(4)$ & $1.140(14)$ & $-0.000158(387)$\\
$5.35$ & $(0.835280)$ & $1500+$ & $0.28$ & $800$ & $200$ & $0.0748(4)$ & $1.138(16)$ & $0.000147(336)$\\
$5.40$ & $(0.837659)$ & $2600+$ & $0.30$ & $1600$ & $400$ & $0.0677(5)$ & $1.120(14)$ & $0.000185(79)$\\
$5.45$ & $(0.839960)$ & $3100+$ & $0.30$ & $2000$ & $400$ & $0.0622(2)$ & $1.060(7)$ & $0.000071(234)$\\
$5.475$ & $(0.841081)$ & $2450+$ & $0.32$ & $1500$ & $300$ & $0.0580(3)$ & $1.063(7)$ & $0.000202(188)$\\
$5.50$ & $0.841873(95)$ & $4100+$ & $0.32$ & $1500$ & $375$ & $0.0570(5)$ & $1.064(10)$ & $0.001716(248)$\\
$5.525$ & $(0.843350)$ & $2600+$ & $0.34$ & $1500$ & $300$ & $0.0536(2)$ & $1.041(7)$ & $0.000205(467)$\\
$5.55$ & $(0.844445)$ & $2600+$ & $0.36$ & $1500$ & $300$ & $0.0508(2)$ & $1.031(3)$ & $0.000545(238)$\\
$5.575$ & $(0.845519)$ & $2450+$ & $0.38$ & $1500$ & $300$ & $0.0483(4)$ & $1.073(17)$ & $0.000956(218)$\\
$5.60$ & $(0.846582)$ & $4100+$ & $0.34$ & $1000\ast $ & $250$ & $0.0456(4)$ & $1.040(12)$ & $0.001615(151)$\\
$5.70$ & $(0.850883)$ & $900+$ & $0.36$ & $400$ & $100$ & $0.0409(1)$ & $0.960(2)$ & $0.002867(470)$\\
\hline\hline
\end{tabular}
\end{center}
\end{table*}

\begin{table*}[th]
\caption{Summary table for $N_f = 8$
with the use of the lattice volume $24^3\times 8$.
{The tadpole factors $u_0$ have been  computed
in ~\cite{Deuzeman:2008sc}.}
}\label{tab:Nf8_Nt8}
\begin{center}
\begin{tabular}{l|llllllll}
\hline\hline
$\betaL$ & $u_0$ & $n_{\mathrm{traj}}$ & $\Delta\tau$ & $n_{\mathrm{ave}}$ & $s_{\mathrm{bin}}$ & $2\PBP$ & $a^2\chi_{\sigma}$ & $\mathrm{Re}~L$\\
\hline
$4.00$ & $(0.793260)$ & $4500$ & $0.12$ & $1000$ & $200$ & $0.1584(8)$ & $2.72(3)$ & $0.00006(113)$\\
$4.10$ & $0.79825$ & $3800$ & $0.12$ & $1200$ & $300$ & $0.1387(9)$ & $2.52(4)$ & $-0.00074(82)$\\
$4.15$ & $0.80175$ & $2400+$ & $0.14$ & $1000$ & $200$ & $0.1319(3)$ & $2.39(1)$ & $0.00005(92)$\\
$4.20$ & $0.8053$ & $6500$ & $0.14$ & $1500$ & $375$ & $0.1232(7)$ & $2.30(5)$ & $-0.00094(85)$\\
$4.25$ & $0.8086$ & $8700+$ & $0.14$ & $8000\ast $ & $2000$ & $0.1069(18)$ & $2.06(15)$ & $0.00584(42)$\\
$4.30$ & $(0.8116)$ & $2570$ & $0.16$ & $1000$ & $200$ & $0.0767(4)$ & $1.63(1)$ & $0.01282(50)$\\
$4.40$ & $0.8192$ & $2570$ & $0.18$ & $1000$ & $200$ & $0.0639(2)$ & $1.48(0)$ & $0.01801(52)$\\
\hline\hline
\end{tabular}
\end{center}
\end{table*}






\end{document}